\documentclass[final]{IEEEtran}
\usepackage{amsthm,amssymb,graphicx,multirow,amsmath,color,amsfonts}%,ulem}
\usepackage{setspace}	% Remove in double column version. Also search for \setstretch in the body of the paper and comment these commands for double column
\usepackage[update,prepend]{epstopdf}
\usepackage[noadjust]{cite}
\usepackage[latin1]{inputenc}
\usepackage{tikz}
\usepackage{pdfpages}
\usepackage{subfigure,array}
\usepackage{tabulary}
\usepackage{multirow}
\usepackage{float}
\usepackage{comment}
\usepackage{mathtools}
\let\subparagraph\relax
\usepackage[compact]{titlesec}
\titlespacing{\section}{0pt}{*1}{*1}
\titlespacing{\subsection}{0pt}{*1}{*1}
\include{notation}
\allowdisplaybreaks % Allows breaking of eqnarray over multiple pages (avoids unnecessary blanks in the document before eqnarray)
\newcommand{\myeq}[1]{\mathrel{\overset{\makebox[0.07pt]{\mbox{(#1)}}}{=}}}

\allowdisplaybreaks

\def\nb0{{\mathbf{0}}}
\def\nb1{{\mathbf{1}}}

\def\nbbE{{\mathbb{E}}}

\def\nbbP{{\mathbb{P}}}

\def\nbbR{{\mathbb{R}}}

\def\nbbZ{{\mathbb{Z}}}

\newtheorem{lemma}{Lemma}

\newtheorem{ndef}{Definition}

\newtheorem{theorem}{Theorem}
\newtheorem{prop}{Proposition}
\newtheorem{cor}{Corollary}

\newtheorem{remark}{Remark}

\def\figref#1{Fig.\,\ref{#1}}%

\def\pc{\mathtt{P_c}}
\def\pr{\mathtt{P_r}}
\def\R{\mathbb{R}}

\def\snr{\mathtt{SNR}}

\def\case{{\sc case}}

\graphicspath{{./Fig/}}
\title{Bandwidth Partitioning and Downlink Analysis in Millimeter Wave  Integrated  Access and Backhaul for 5G}

\author{
Chiranjib Saha, Mehrnaz Afshang, and Harpreet S. Dhillon 
\thanks{
C. Saha and H. S. Dhillon are with Wireless@VT, Department of
ECE, Virginia Tech, Blacksburg, VA 24061 USA (e-mail: csaha@vt.edu;
hdhillon@vt.edu).
M. Afshang was with Wireless@VT, Department of ECE, Virginia Tech,
Blacksburg, VA 24061 USA. She is now with Ericsson Research, Santa Clara,
CA 95054 USA (e-mail: mehrnaz@vt.edu). 
This paper was  presented in part at the IEEE International Conference on Communications (ICC), 2018~\cite{saha2017integrated}.
Manuscript last updated: \today.
} }

\let\emptyset\varnothing
\begin{document}
\maketitle
\begin{abstract}
With the increasing  network densification, it has
 become exceedingly difficult to provide traditional fiber backhaul
 access to each cell site, which is especially true for small cell
 base stations (SBSs). The increasing maturity of millimeter wave
 (mm-wave) communication has opened up the possibility of
 providing high-speed wireless backhaul to such cell sites. Since
mm-wave is also suitable for access links, the third generation
 partnership project (3GPP) is envisioning an integrated access
 and backhaul (IAB) architecture for the fifth generation (5G)
 cellular networks in which the same infrastructure and spectral
 resources will be used for both access and backhaul. In this paper,
 we develop an analytical framework for IAB-enabled cellular
 network using which its downlink rate coverage probability
 is accurately characterized. Using this framework, we study
 the performance of three backhaul bandwidth (BW) partition
 strategies: 1) equal partition: when all SBSs obtain equal share of
 the backhaul BW; 2) instantaneous load-based partition: when the
 backhaul BW share of an SBS is proportional to its instantaneous
 load; and 3) average load-based partition: when the backhaul BW
 share of an SBS is proportional to its average load. Our analysis
 shows that depending on the choice of the partition strategy,
 there exists an optimal split of access and backhaul BW for
 which the rate coverage is maximized. Further, there exists a
 critical volume of cell-load (total number of users) beyond which
 the gains provided by the IAB-enabled network disappear and
 its performance converges to that of the traditional macro-only
 network with no SBSs.
\end{abstract}
\section{Introduction}\label{sec::intro}
With the exponential rise in data-demand far exceeding the capacity of the traditional macro-only cellular network operating in sub-6 GHz bands, network densification using   mm-wave base stations (BSs) is becoming a major driving technology for the 5G wireless evolution~\cite{dehos2014millimeter}. While heterogeneous cellular networks (HetNets) with low power SBSs overlaid with traditional macro BSs  improve the spectral efficiency  of the access link (the link between a user and its serving BS), mm-wave communication can further boost the data rate by offering high bandwidth. {That said, one of the main hindrances in the way of large-scale deployment of small cells is that}  the existing high-speed optical fiber backhaul network that connects the BSs to the network core is not scalabale to the extent of ultra-densification envisioned for  small cells~\cite{quek2013small,tipmongkolsilp2011evolution,DhillonCaireBackhaul}. However, with recent advancement in mm-wave communication with highly directional beamforming~\cite{rangan2014millimeter,GhoshMMwave}, it is possible to   replace the so-called {\em last-mile fibers} for  SBSs by establishing fixed  mm-wave backhaul links  between the SBS and the MBS  equipped with fiber backhaul, also known as the anchored BS (ABS), thereby achieving Gigabits per second (Gbps) range data rate over backhaul links~\cite{GaoMassiveMimomm-waveBackhaul}. While mm-wave fixed wireless backhaul is targetted to be  a part of the first phase of the commercial roll-out   of 5G~\cite{mm-waveMagazineDahlmamn},  3GPP is exploring a more ambitious solution of IAB  
where the ABSs will use {the}  same spectral  resources and infrastructure of mm-wave transmission to serve  cellular users in access as well as the SBSs in backhaul~\cite{accessbackhaul3gpp}. In this paper,  we develop a  tractable analytical framework {for} IAB-enabled mm-wave cellular networks using tools from stochastic geometry and obtain some design insights that will be useful for the ongoing pre-deployment studies on IAB. %more extensive studies by 3GPP as well as the future deployment of IAB-enabled cellular networks. 

 \subsection{Background and related works}
Over recent years, stochastic geometry has emerged as a powerful tool for modeling and analysis of cellular networks operating in sub-6 GHz~\cite{haenggi2012stochastic}. 
  The locations of the BSs and users are commonly modeled  as independent Poisson point processes (PPPs) over an infinite plane.  
This model,  initially developed for the analysis of traditional macro-only cellular networks~\cite{AndrewsTractable},  was further extended for  the analysis of HetNets in~\cite{dhillon2012modeling,mukherjee2012distribution,Prasanna_globecomm,jo2012heterogeneous}. 
In the followup works, this PPP-based HetNet model was used to study many different aspects of cellular networks such as load balancing, BS cooperation, multiple-input multiple-output (MIMO), energy harvesting, and many more. Given the activity this area has seen over the past few areas, any attempt towards summarizing all key relevant prior works here would be futile. Instead, it would be far more beneficial for the interested readers to refer to dedicated surveys and tutorials~\cite{elsawy2013stochastic,elsawy2016modeling,andrews2016primer,mukherjee2014analytical}  that already exist on this topic. While these initial works were implicitly done for cellular networks operating in the sub-6 GHz spectrum, tools from stochastic geometry have also been leveraged further to characterize their performance in the mm-wave spectrum~\cite{AndrewsMMWave,Direnzo_mmwave,MillimeterWaveBai2015,mmWaveHetNetTurgut}.
%
%
%
%
%These early works enabled the {analytical performance evaluation} of different  technologies of current and future cellular networks, such as multiple-input-multiple output (MIMO), load balancing, energy harvesting  BS-cooperation, and many more.  Since this line of work is \chb{quite  well-known} by now, we do not provide a comprehensive discussion on the prior arts and refer the interested readers to the dedicated surveys and tutorials  
%\cite{elsawy2013stochastic,elsawy2016modeling,andrews2016primer,mukherjee2014analytical} for a more pedagogical treatment
%of this general research direction.   With the emergence of mm-wave communication as a {viable} solution to alleviate the spectrum crunch in  sub-6 GHz, the tools from stochastic geometry have been further leveraged to study mm-wave cellular networks~\cite{AndrewsMMWave,MillimeterWaveBai2015,mm-waveHetNetTurgut}. 
These mm-wave cellular network models specifically focus on the  mm-wave propagation characteristics which signifantly differ from those of the sub-6 GHz~\cite{rangan2014millimeter}, such as the severity of blocking of mm-wave signals  by physical obstacles like walls and trees, directional beamforming using antenna arrays, and inteference being dominated by noise~\cite{Kulkarni_backhaul_asilomar}.  % Taking these factors into consideration, the PPP-based mm-wave cellular network model was proposed in~\cite{MillimeterWaveBai2015,mm-waveHetNetTurgut}. 
 These initial modeling approaches were later extended to  study different problems specific to mm-wave cellular networks, such as,  
 cell search~\cite{liAndrews2017directional}, antenna beam alignment~\cite{HeathAlkhateeb2017BeamAssociation}, {and cell association in the mm-wave spectrum~\cite{Elshaer_cell_association}.}  With this brief introduction, we now shift our attention to the main focus of this paper which is IAB in mm-wave cellular networks. In what follows, we provide the rationale behind mm-wave IAB   and how stochastic geometry can {be used for its performance evaluation.}

 For traditional cellular networks, it is  reasonable to assume that the capacity achieved by the  access links is not limited by the backhaul constraint on the serving BS  since  all BSs have access to the high capacity wired backhaul. As expected, backhaul constraint was ignored in almost all prior works on stochastic geometry-based modeling and analysis of cellular networks. 
 However, with the increasing  network densification with small cells, it may not be feasible to connect every SBS to  the wired backhaul network which is limited by cost, infrastructure, maintenance, and  scalability. These limitations motivated a significant body of research works %in last decade 
on the expansion of the cellular networks by deploying  relay nodes connected to the ABS by  wireless backhaul links, {e.g. see}~\cite{backhaul_survey_1}. 
Among different techniques of wireless backhaul, 3GPP included layer 3 relaying as a part of the long term evolution advanced (LTE-A) standard in Release 10~\cite{relay3gpp1,relay3gpp2} for coverage extension of the cellular network. Layer 3 relaying follows the principle of IAB architecture, which is often synonymously referred to as {\em self-backhauling}, where the relay nodes have the functionality of SBS and the ABS multiplexes  its time-frequency resources to establish access links with the users and wireless backhaul links with SBSs that may not have access to wired backhaul~\cite{johansson2016self}. However, despite being the part of the standard, %SBSs with wireless backhaul 
 layer 3 relays have never  really been deployed on a massive scale in 4G mostly due to  the spectrum shortage in sub-6 GHz. For instance, in urban regions with high capacity demands, the operators are not willing to {relinquish} {any part  of the  cellular  bandwidth (costly and scarce resource)} for wireless backhaul. However, with recent advancement in mm-wave communication, IAB has gained substantial interest since spectral bottleneck will not be a primary concern once high bandwidth in mm-wave spectrum (at least 10x the cellular BW in sub-6 GHz) is exploited. Some of the notable industry initiatives driving mm-wave IAB are  mm-wave small cell access and backhauling
(MiWaveS)~\cite{miWaveS} and 5G-Crosshaul~\cite{crosshaul}. In 2017, 3GPP also {started working on} a new study item to investigate the performance of IAB-enabled mm-wave cellular network~\cite{accessbackhaul3gpp}.
  
Although backhaul is becoming a primary bottleneck of cellular networks, there is  very little existing work on the stochastic geometry-based analyses considering the backhaul constraint~\cite{QuekBackhaul,suryaprakash2014analysis,SinghAndrews2014}. 
 While these works are focused on the traditional networks in sub-6 GHz, contributions on  IAB-enabled mm-wave HetNet are even sparser, except an extension of the PPP-based model~\cite{SinghKulkarniSelfBackhaul}, where the authors modeled wired and wirelessly backhauled BSs and users as three independent PPPs. In \cite{Ganti-self-backhaul,tabassum2016analysis}, similar modeling approach was used to study IAB in sub-6 GHz using full duplex BSs.   
 The fundamental shortcoming of these PPP-based models is the assumption of independent locations of the BSs and users which are spatially coupled in actual networks. For instance, in reality, the users form spatial clusters, commonly known as {\em user hotspots} and   the centers of the user hotspots are targetted  as the potential  cell-cites of the short-range mm-wave SBSs~\cite{Saha_J1}.  Not surprisingly, such spatial configurations of users and BSs are at the heart of the 3GPP simulation models~\cite{saha20173gpp}.  
 To address this shortcoming of the analytical models,  
in this paper, we propose the {\em first 3GPP-inspired} {\em stochastic geometry-based} finite network model for the performance analysis of HetNets with IAB.  The key contributions are summarized next. 

\subsection{Contributions and outcomes} \label{subsec::contributions}
\subsubsection{New tractable model for IAB-enabled mm-wave HetNet}
We develop a realistic and tractable analytical framework to study the performance of IAB-enabled mm-wave HetNets. % where the ABS serves users in access links and SBSs in backhaul links from the same pool of spectral resources.  expand on the model
Similar to the models used in 3GPP-compliant {simulations}~\cite{accessbackhaul3gpp}, we consider a two-tier HetNet where a  {circular macrocell with ABS at the center} is overlaid by numerous low-power small cells.   The users are assumed to be non-uniformly distributed over the macrocell forming hotspots and the SBSs are located at the geographical centers of these user hotspots. {The non-uniform distribution of the users and the spatial coupling of their locations with those of the SBSs means that the analysis of this setup is drastically different from the state-of-the-art PPP-based models. Further, the consideration of a single macrocell (justified by the noise-limited nature of mm-wave communications), allows us to glean crisp insights into the coverage zones, which further facilitate a novel analysis of load on ABS and SBSs\footnote{In our discussion, BS load refers to the number of users connected to the BS.}.} Assuming that the total system BW is partitioned into two splits for access and backhaul communication,  we use this model to study the performance of three backhaul  BW partition strategies, namely, (i) {\em equal partition}, where each SBS gets equal share of  BW irrespective of its load, (ii) {\em instantaneous load-based partition}, where the ABS frequently collects information from the SBSs on  their  instantaneous loads and partitions the backhaul BW  proportional to the {instantaneous} load on each SBS, and (iii) {\em average load-based partition}, where the ABS  collects information from the SBSs on {their average loads} and partitions the backhaul BW  proportional to the average load on each SBS. 
\subsubsection{New load modeling and downlink rate analysis}
For the purpose of performance evaluation and comparisons between the aforementioned strategies, we evaluate the downlink rate coverage probability i.e. probability that the downlink data rate experienced by a randomly selected user will exceed a target data rate.
% assuming that  Shannon rate is achieved between any two communicating nodes.  
As key intermediate steps of our analysis we  {characterize} the two essential components of  rate coverage, which are (i) {signal-to-noise-ratio} ($\snr$)-coverage probability, and (ii) the distribution of ABS and SBS load, {which} directly impacts the amount of resources allocated by the serving BS to the user of interest. 
%$\snr$ appears inside the $\log$ term and load appears as a pre-log factor in the Shannon capacity
%formula. 
We compute the {probability mass functions} (PMFs) of the ABS and the SBS loads assuming the number of users per hotspot is  fixed.  We {then} relax this fixed user assumption by considering independent Poisson distribution on the number of users in each hotspot. Due to a significantly different spatial model, our approach of load modeling  is quite different from the load-modeling in PPP-based networks~\cite{SinghAndrews2014}.  
\subsubsection{System design insights} 
Using the proposed analytical framework, we obtain the following system design insights. 
\begin{itemize}
\item  We {compare} the three backhaul BW partition strategies in terms of three metrics, (i) rate coverage probability, (ii)  median rate, and  (iii) $5^{th}$ percentile rate.  {Our numerical results indicate  that for a given  combination of the backhaul BW partition strategy and the performance metric of interest, there exists an  optimal access-backhaul
BW split for which the metric is maximized.}  
\item Our  results  demonstrate that the optimal access-backhaul partition fractions for median and $5^{th}$ percentile rates are not  very sensitive to the choice of backhaul BW partition strategies. 
%While the optimal $5^{th}$ percentile rate is very sensitive to the choice of the \chb{access-backhaul partition fraction}, the median  rate is relatively flat around the optimal partition fraction.
 Further, the median and $5^{th}$ percentile rates are invariant to  system BW.  
%\item In terms of the  rate coverage under optimal access-backhhaul BW split, the instantaneous  load-based partition dominates the average load-based partition which further dominates the equal partition. 
%\item The assumption of the number of users being fixed or Poisson distributed does not significantly impact the performance trend in terms of rate coverage probability. \chb{This insight may be useful for future studies on  IAB-enabled networks where the assumption of fixed number of users per hotspot can significantly reduce the  complexity of the system model.} 
\item For  given infrastructure and spectral resources, the IAB-enabled network  outperforms the macro-only network with no SBSs up to a critical volume of total cell-load, beyond which the  performance gains disappear and its performance  converges to that of the macro-only network. Our numerical results also indicate that this critical total  {cell-load} increases almost linearly with the system BW.  
%\item We have compared the three backhaul BW partition strategies in terms of other metrics such as median and $5^{th}$ percentile rates.  While the optimal $5^{th}$ percentile rate is very sensitive to the choice of the \chb{access-backhaul partition fraction}, the median  rate is relatively flat around the optimal partition fraction. %For median and  $5^{th}$
 \end{itemize}
\section{System Model}\label{sec::system::model}
 \subsection{mm-wave Cellular System Model}
%\begin{minipage}{\linewidth}
%      \centering
\begin{figure}
          \centering
          \subfigure[User and BS locations]{
              \includegraphics[width=.98\linewidth]{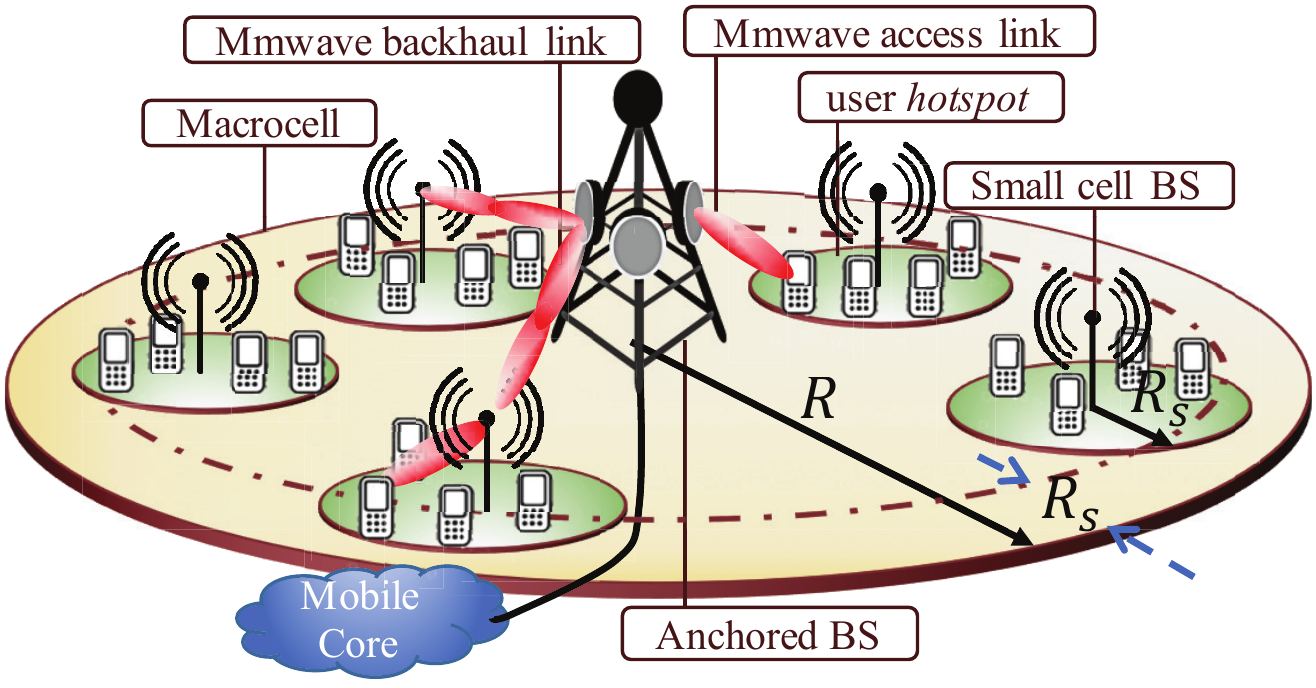}
             \label{fig::system::model::1}}
              \subfigure[Resource Allocation]{
               \label{fig::system::model::2}
              \includegraphics[width=.99\linewidth]{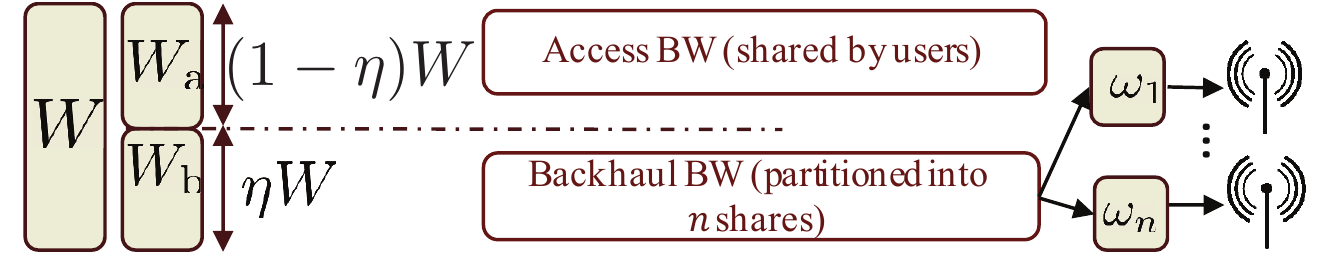}
              }
              \caption{Illustration of the system model.}
             \label{fig::system::model}
          \end{figure}   
%\end{minipage}
\subsubsection{BS and user locations} Inspired by the spatial configurations used in 3GPP {simulations}~\cite{accessbackhaul3gpp,saha20173gpp} for a typical outdoor deployment scenario of a two-tier HetNet, we assume  that $n$ SBSs are deployed inside a  circular macrocell of radius $R$ (denoted by $b({\bf 0},R)$) with the macro BS at its center. We assume that this BS is  connected to the core network with high speed optical fiber and  {is hence} an ABS. Note that, in contrast to  the  infinite network models (e.g. the PPP-based networks defined over $\R^2$) which are suitable for interference-dominated networks (such as conventional cellular networks in sub-6 GHz), we are limiting the complexity of the system model by considering single macrocell. This assumption is  justified by the noise-limited nature of mm-wave communications~\cite{Kulkarni_backhaul_asilomar}.   {Moreover, as will be evident in the sequel, this setup will allow us to glean crisp insights into the properties of this network despite a more general user distribution model (discussed next) compared to the PPP-based model.} 

We model a user hotspot at ${\bf x}$ as  $b({\bf x},R_{\rm s})$, i.e., a circle of radius $R_{\rm s}$ centered at ${\bf x}$. We assume that the macrocell contains $n$ user hotspots,  located at $\{{\bf x}_i\equiv (x_i,\varphi_i), i=1,\dots,n\}$, which are distributed uniformly at random in $b({\bf 0},R-R_{\rm s})$.\footnote{{For notational simplicity, we use $x\equiv\|{\bf x}\|,\ \forall\ {\bf x}\in\nbbR^2$.}} Thus, $\{{\bf x}_i\}$ is a sequence of independently and identically distributed (i.i.d.) random vectors with the  distribution of ${\bf x}_i$ being:
\begin{align}\label{eq::sbs::distribution}
f_{\bf X}({\bf x}_i)=\begin{cases}\frac{x_i}{\pi(R-R_{\rm s})^2}, &\text{when }0<x_i\leq R- R_{\rm s}, 0<\varphi_i\leq 2\pi,\\
0, &\text{otherwise.}
\end{cases}
\end{align}The marginal probability density function (PDF) of $x_i$ is obtained as: $f_{X}(x_i)=2x_i/(R-R_{\rm s})^2$ for $0<x_i\leq R- R_{\rm s}$ and $\varphi_{i}$ is a uniform random variable in $(0,2\pi]$.
Note that this construction  ensures that all hotspots lie entirely  inside the macrocell, i.e., $b({\bf x}_i,R_{\rm s})\cap b({\bf 0},R)^c = \emptyset,\ \forall\ i$. 
%From this point, we will frequently call a hotspot centered at ${\bf x}$ as hotspot at ${\bf x}$.  
 {We  assume that the number of users in the hotspot centered at ${\bf x}_i$ is $N_{{\bf x}_i}$, where} \case~$1$: $N_{{\bf x}_i}=\bar{m}$ is  fixed and equal for all $i=1,\dots, n$ and \case~$2$: $\{N_{{\bf x}_i}\}$ is a sequence of i.i.d. Poisson random variables with mean $\bar{m}$.
%The analysis for the case  where number of users per cluster is independently and identically distributed (i.i.d.) Poisson random variable is deferred to an extended version of this paper.  
These ${N}_{{\bf x}_i}$ users are assumed to be located {\em uniformly at random} {independently of each other}  in each hotspot. Thus, the location of a user belonging to the hotspot at ${\bf x}_i$ is denoted by ${\bf x}_i+{\bf u}$, where ${\bf u} \equiv (u,\xi)$ is a random vector in $\R^2$ with PDF: 
\begin{align}\label{eq::users::distribution}
f_{\bf U}({\bf u})=\begin{cases}\frac{u}{\pi R_{\rm s}^2}, &\text{when }0< {u}\leq R_{\rm s}, 0<\xi\leq 2\pi\\
0, &\text{otherwise.}
\end{cases}
\end{align} {The marginal PDF of $u$} is: $f_{U}(u)=2u/R_{\rm s}^2$ for $0<u\leq R_{\rm s}$ and $\xi$ is a uniform random variable in $(0,2\pi]$.    We assume that the SBSs are deployed at the center of user hotspots, i.e., at $\{{\bf x}_i\}$. The ABS provides wireless backhaul to these SBSs over  mm-wave links. See Fig.~\ref{fig::system::model::1} for an illustration. Having defined the spatial distribution of SBSs and users, we now define the {\em typical user} for which we will compute the rate coverage probability. The typical user is  a user chosen uniformly at random from the network. The hotspot to which the typical user belongs is termed as the {\em representative hotspot}.   We denote the center of representative hotspot as   ${\bf x}$, where ${\bf x}={\bf x}_n$, without loss of generality and the location of the typical user as ${\bf x}+{\bf u}$.  For \case~$1$, the number of users in the representative cluster is $N_{\bf x}=N_{{\bf x}_n}=\bar{m}$.   For \case~$2$, although $N_{{\bf x}_i}$ is i.i.d. Poisson, $N_{\bf x}$  {does not follow the same distribution} since the typical user will more likely belong to a hotspot with more number of users~\cite{Qin2017}. 
If $n\to \infty$, $N_{\bf x}$  
follows a weighted Poisson distribution with PMF  $
\nbbP(N_{\bf x} = k ) = \frac{\bar{m}^{k-1}e^{-\bar{m}}}{(k-1)!}$, where,  $k\in\nbbZ^+$. It can be easily shown that if $N_{\bf x}$ follows a weighted Poisson distribution, {we have} $N_{\bf x} = N_{{\bf x}_n}+1$. Hence, for $n\to \infty$, one can obtain the  distribution of $N_{\bf x}$ by first choosing  {a} hotspot uniformly at random and then   {adding one user to it}. However, when $n$ is finite, $N_{\bf x}$ 
 will lie between $N_{{\bf x}_{n}}$ and $N_{{\bf x}_{n}}+1$ ($N_{{\bf x}_n}\leq N_{\bf x}\leq N_{{\bf x}_{n}}+1$). The lower bound on $N_{\bf x}$ is trivially achieved when $n=1$. %since this is equivalent to sampling from a Poisson distribution.   
 Since the actual distribution of $N_{\bf x}$ for finite number of hotspots ($n>1$) is not tractable, we fix the typical user for \case~$2$ according to the following Remark. 
\begin{remark}\label{rem::typical} For \case~$2$, 
we first  choose a  hotspot centered at ${\bf x}$ uniformly at random from $n$ hotspots, call it the representative hotspot,  and then add  the typical user at ${\bf x}+{\bf u}$, where $\bf u$ follows the PDF in \eqref{eq::users::distribution}. Although this process of selecting the typical user is asymptotically exact when $N_{{\bf x}_i}\stackrel{i.i.d.}{\sim}{\tt Poisson}(\bar{m}),\ \forall\ i = 1,2,\dots,n$, and $n\to\infty$,    
 it will have negligible impact on the  analysis since {our interest will be in the cases where the macrocells have moderate}  to high number of hotspots~\cite{3gppreportr12}. 
\end{remark}

%Typical user description needs modification. Here is one possible order: Say that typical user is a user chosen uniformly at random from the network. Call the cluster to which it belongs as the representative cluster. Then say that in case I, the representative cluster will have N_{x_n} users. However, in case 2, typical user is more likely to be from the cluster with more points. Hence, it is reasonable to say that the representative cluster will have more than N_{x_n} users. If the number of clusters would have been infinite, this would have been wighted Poisson distribution, which essentially means that the number of users in this cluster are N_{x_n}+1. However, since we have finite clusters, the actual number of users lie between N_{n_n} and N_{x_n}+1. Since a macrocell will have moderate to large number of clusters, we use N_{x_n}+1 points .......

\subsubsection{Propagation assumptions}\label{subsec::prop::assumption}
 All backhaul and access transmissions are assumed to be performed in mm-wave spectrum. {We assume that the ABS and SBS transmit at  constant power spectral densities (PSDs) $P_{\rm m}/W$ and $P_{\rm s}/W$, respectively over  a system BW $W$.   The received power at ${\bf z}$ is given by $P \psi h L({\bf z},{\bf y})^{-1}$, where $P$ is a generic variable denoting transmit power with $P\in\{P_{\rm m},P_{\rm s}\}$, $\psi$ is the combined antenna gain of the transmitter and receiver, and $L ({\bf z},{\bf y})= 10^{((\beta +10\alpha\log_{10}\|{\bf z}-{\bf y}\|)/10)}$ is the associated pathloss.}  We assume that all links undergo i.i.d. Nakagami-$m$ fading. Thus, $h\sim{\tt Gamma}(m,m^{-1})$.
%(\chb{shadowing should be ignored as we can not apply displacement theorem in this context.})
\subsubsection{Blockage model}
\label{subsec::blockagemodel}
%Each access link of separation $d$ is assumed to be LOS with probability $C$ if $d\leq D$ and $0$ otherwise. The ordered pair $(C,D)$ differ depending on the geography and deployment scenario (for instance, low for dense urban, high for semi-urban). 
%
%\chb{Instead of LOS ball model, the exponential model is better tractable for our setup (see Heath beamforming paper). }
Since mm-wave signals are sensitive to physical blockages such as buildings, trees and even human bodies,   the LOS and NLOS path-loss characteristics have to be explicitly included into the analysis.   
On similar lines of \cite{Bai_mmWave}, we assume exponential blocking model.  Each mm-wave link of distance $r$ between the transmitter (ABS/SBS) and receiver (SBS/user) is LOS or NLOS according to an independent Bernoulli random variable with LOS probability $p(r) =\exp(-r/\mu)$, where $\mu$ is the LOS range constant that depends on the geometry and density of blockages. {Since the blockage environment seen by the links between the ABS and SBS, SBS to user and ABS to user may be very different, one can assume three different blocking constants $\{\mu_{\rm b},\mu_{\rm s},\mu_{\rm m}\}$, respectively instead of  a single blocking constant $\mu$. As will be evident in the technical exposition, this does not require any major changes in the analysis. However, in order to keep our notational simple, we will assume the same $\mu$ for all the links in this paper.} Also, LOS and NLOS links may likely follow  different fading statistics, which is incorporated by assuming different  Nakagami-$m$ parameters for LOS and NLOS, denoted by $m_L$ and $m_{NL}$, respectively.

We assume that all BSs are equipped with steerable directional antennas and the user equipments have omni-directional antenna. Let $G$ {be} the directivity gain of the transmitting and receiving antennas of the BSs (ABS and SBS). Assuming perfect beam alignment, the effective gains on  backhaul and  access links are  $G^2$ and $G$, respectively. We assume that the system is noise-limited, i.e., at any receiver, the  interference is negligible compared to the thermal noise  with PSD ${\tt N}_0$.   
%Noticing that the  $\snr$ on a link with BW ${W}'$  is proportional to $\frac{\frac{P}{W}{W}'}{{\tt N}_0{W}'} = P/{\tt N}_0W$, 
Hence, the $\snr$-s of a backhaul link   from ABS to SBS at ${\bf x}$,  access links from SBS at ${\bf x}$ to user at ${\bf x}+{\bf u}$, and ABS to user at ${\bf x}+{\bf u}$  are respectively expressed as:
\begin{subequations}\label{eq::sbr::equations}
\begin{alignat}{3}
&\snr_{\rm b}({\bf x}) = \frac{P_{\rm m} G^2 h_{\rm b}L({\bf 0},{\bf x})^{-1}}{{\tt N}_0 W},\\ 
&\snr_{\rm a}^{\rm SBS}({\bf x}+{\bf u}) =  \frac{P_{\rm s}G h_{\rm s}L({\bf x},{\bf x}+{\bf u})^{-1}}{{\tt N}_0W},\\
&\snr_{\rm a}^{\rm ABS}({\bf x}+{\bf u}) =  \frac{P_{\rm m}
G h_{\rm m}L({\bf 0},{\bf x}+{\bf u})^{-1}}{{\tt N}_0W},
\end{alignat}
\end{subequations}
where $\{h_{\rm b}, h_{\rm s}, h_{\rm m}\}$ are the corresponding small-scale fading gains.% associated with the links. %, We assume each link undergoes i.i.d. Nakagami-$m$ fading, i.e., $h_{\rm b}$, $h_{\rm s}$, $h_{\rm m}\sim {\tt Gamma}(m,m^{-1})$.
\subsubsection{User association}\label{subsec::user::association}
{ We assume that the SBSs operate in closed-access, i.e., users in hotspot can only connect to the SBS at {the} hotspot center, or the ABS.  This model is inspired by the way smallcells with closed user groups, for instance the privately owned femtocells, are dropped in  the HetNet models considered by 3GPP~\cite[Table A.2.1.1.2-1]{access2010further}.} 
Given the complexity of user association in mm-wave using beam sweeping techniques, we assume a much simpler way of  user association which is performed by signaling in sub-6 GHz, analogous to the current LTE standard~\cite{HeathAlkhateeb2017BeamAssociation}. In particular, the BSs broadcast paging signal using omni-directional antennas in sub-6 GHz and the user associates to the candidate serving BS based on the maximum received power over the paging signals.  
 Since the broadcast signaling is in sub-6 GHz, we   assume the  same power-law pathloss function for both LOS and NLOS components  with path-loss exponent $\alpha$ due to rich scattering environment. 
 We define the association event  ${\cal E}$  for the typical user as:
 \begin{align}
 {\cal E} = \begin{cases}
 1 &\text{ if } P_{\rm s}\|{\bf u}\|^{-\alpha} >P_{\rm m}\|{\bf x}+{\bf u}\|^{-\alpha}, \\
  0, &\text{ otherwise,}
 \end{cases}
 \end{align}
 where $\{0,1\}$ denote association to ABS and SBS,  respectively. The typical user at ${\bf x}+{\bf u}$ is  {\em under coverage} in the downlink if either of the following two events occurs:
 \begin{align}
  & {\cal E}  = 1 \text{ and }  \snr_{\rm b}({\bf x})>\theta_1, \snr_{\rm a}^{\rm SBS}({\bf u})>\theta_2, \text{ or,}\notag\\
  &{\cal E}  = 0 \text{ and }\snr_{\rm a}^{\rm ABS}({\bf x}+{\bf u})>\theta_3,\label{eq::coverage::def}
 \end{align}
 where $\{\theta_1, \theta_2, \theta_3\}$ are  the coverage thresholds for successful demodulation and decoding. 
\subsection{Resource allocation} \label{subsec::resourceallocation}
{The ABS,  SBSs and users are assumed to be capable of communicating on both mm-wave and sub-6 GHz bands.}  The sub-6 GHz band is reserved for control channel and the mm-wave band is kept for data-channels. 
%Since mm-wave links are unreliable for long distances, a fraction of the macro users  may be offloaded to sub-6 GHz depending on the  link quality from ABS to the user. However, due to space constraints,  we will delegate the discussion on sub-6 GHz offloading to the extended version of this paper.   
The total mm-wave BW $W$ for downlink transmission is partitioned into two parts, $W_{\rm b}=\eta W$ for backhaul and $W_{\rm a}=(1-\eta)W$ for access, where $\eta\in[0,1)$ determines the access-backhaul split. Each BS is assumed to employ a simple round robin scheduling policy for serving users, under which the total access BW is {shared equally}  among its associated users, referred to  alternatively as {\em load} on that particular BS. 
  On the other hand,  the backhaul BW is shared amongst $n$ SBSs by  either of the three strategies as follows. \begin{enumerate}
\item {\em Equal partition.} This is the  simplest partition strategy  where the ABS does not require any load information from the SBSs and divides $W_{\rm b}$  equally  into $n$  splits. 
\item {\em Instantaneous load-based partition.} In this scheme, the SBSs regularly feed back the ABS its load information and accordingly the  ABS allocates backhaul BW proportional to the {instantaneous}  load on each small cell.
\item   {\em Average load-based partition.} Similar to the previous strategy, the  ABS allocates backhaul BW proportional to the  load on each small cell. But in this scheme, the SBSs  feed back the ABS its load information after {sufficiently} long intervals.    Hence the instantaneous fluctuations in SBS load  are averaged out.
  \end{enumerate}
    If  the SBS at ${\bf x}$ gets backhaul BW  $W_{\rm s}({\bf x})$, then 
\begin{align}\label{eq::bandwidth::partition}
W_{\rm s}({\bf x}) =
% \left\{\begin{array}{@{}ll@{}} 
\begin{cases}
\frac{W_{\rm b}}{n}, & \text{for equal partition},\\
\multirow{2}{*}{${\frac{N^{\rm SBS}_{{\bf x}}}{N^{\rm SBS}_{{\bf x}}+\sum\limits_{i=1}^{n-1}N^{\rm SBS}_{{\bf x}_i}}} W_{\rm b}$,}& \text{for instantaneous }\\&\text{load-based partition},\\
\multirow{2}{*}{$\frac{\nbbE[N^{\rm SBS}_{{\bf x}}]}{\nbbE[N^{\rm SBS}_{{\bf x}}]+\sum\limits_{i=1}^{n-1}\nbbE[N^{\rm SBS}_{{\bf x}_i}]} W_{\rm b}$,}& \text{for average load-based}\\&\text{ partition},
%\end{array}\right.
\end{cases}
\end{align}% irrespective of their load. 
where $N_{\bf x}^{\rm SBS}$ and $N_{{\bf x}_i}^{\rm SBS}$  denote the load on the SBS of the representative hotspot and load on the  SBS at ${\bf x}_i$, respectively. The BW partition is illustrated in Fig.~\ref{fig::system::model::2}.

%We also similarly define $N_{\rm z}^{\rm ABS}$  the load on the ABS due to the users in hotspot at ${\bf z}$. 
 %Then,  the SBS at ${\bf x}$ is allocated backhaul bandwidth, 
%The distributions of $N^{\rm SBS}_{{\bf x}}$ and $N^{\rm SBS}_{\rm o}$ will be discussed in Section~\ref{subsec::load::dist}. 
% Note that, from implementation complexity, equal partition is the simplest partition policy as the ABS blindly divides the BW in $n$ equal shares. However, the overhead of the information being fed back from the SBS to ABS increases with the frequency of update and  becomes worst for the instantaneous load-based partition.
  To compare the performance of these strategies, we  define the network performance metric of interest next. 
\subsection{Downlink data rate}\label{subsec::downlink::data::rate}
The maximum achievable  downlink data rate, henceforth referred to as simply the {\em data rate}, on the backhaul link between the ABS and the SBS, the access link between SBS and user, and the access link between ABS and user can be expressed as:
\begin{subequations}
\begin{alignat}{3}
{\cal R}_{\rm b}^{\rm ABS} &=  W_{\rm s}({\bf x})
\log_2(1+\snr_{\rm b}({\bf x})),\label{eq::rate_backhaul}\\
{\cal R}_{\rm a}^{\rm SBS} &= \min\bigg(\frac{W_{\rm a}}{N_{\bf x}^{\rm SBS}}\log_2(1+\snr_{\rm a}^{\rm SBS}({\bf u})),
\frac{ {\cal R}_{\rm b}^{\rm ABS}}{N_{\bf x}^{\rm SBS}}\bigg),\label{eq::rate_sbs_access} \\
{\cal R}_{\rm a}^{\rm ABS} &= \frac{W_{\rm a} }{N_{\bf x}^{\rm ABS}+\sum\limits_{i=1}^{n-1}N_{{\bf x}_i}^{\rm ABS}}\log_2(1+\snr_{\rm a}^{\rm ABS}({\bf x}+{\bf u})),
%,\notag\\ &\qquad\qquad\qquad\qquad\text{ if } \snr_{\rm a}%^{\rm SBS}({\bf x}+{\bf u})>\theta_{\rm m},
\label{eq::rate_abs_access}
\end{alignat}\label{eq::rate}
\end{subequations}
where ${W_{\rm s}}({\bf x})$ is defined  according to backhaul BW  partition strategies in \eqref{eq::bandwidth::partition} and  $N_{\bf x}^{\rm ABS}$ ($N_{{\bf x}_i}^{\rm ABS}$) denotes the load on the ABS due to the macro users of the representative hotspot (hotspot at ${\bf x}_i$). In \eqref{eq::rate_sbs_access}, the first  term inside the $\min$-operation is the data rate achieved under no backhaul constraint  when the access BW $W_{\rm a}$ is equally partitioned between $N_{\bf x}^{\rm SBS}$ users. However, due to finite backhaul,  ${\cal R}_{\rm a}^{\rm SBS}$ is limited by the second term. 
%\chb{An implicit assumption while writing \eqref{eq::rate_backhaul} is the fact that the SBSs are operating in half-duplex mode.} 
%When the user is served by the ABS, we assumed that the access link can be either in mm-wave (if $\snr({\bf x}+{\bf u})>\theta_{\rm m}$), or in sub 6 GHz. 
% In this paper, we are only interested in the rate-coverage in mm-wave and hence the users offloaded to sub-6 GHz is outside the purview of our analysis.  
\section{Rate Coverage Probability Analysis}
In this Section, we derive the expression of  rate coverage probability of the typical user  conditioned on its location at ${\bf x}+{\bf u}$ and later decondition over them. This deconditioning step averages out all the spatial randomness of the user and hotspot locations in the given network configuration. We first partition  each hotspot into  SBS and ABS association regions such that the users lying in the SBS (ABS) association region  connects to the SBS (ABS). Note that the formation  of these {mathematically tractable} association regions is the basis of the distance-dependent load modeling which is one of the major contributions of this work.
\begin{figure}
\centering
\includegraphics[scale=0.48]{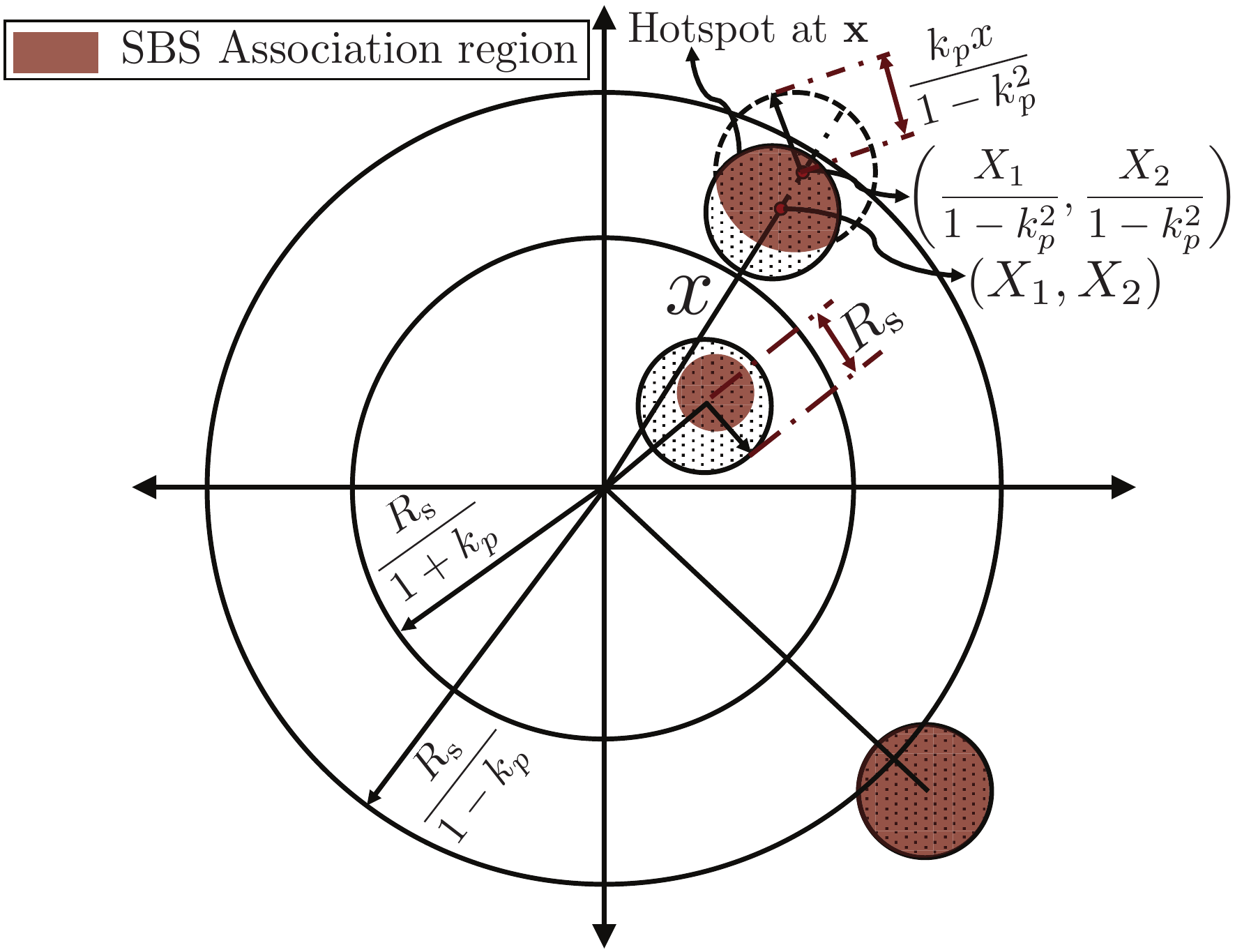}
\caption{An illustration of association region.}
\label{fig::association::construction}
\end{figure}
\begin{figure}
\centering
\includegraphics[width = 0.8\linewidth]{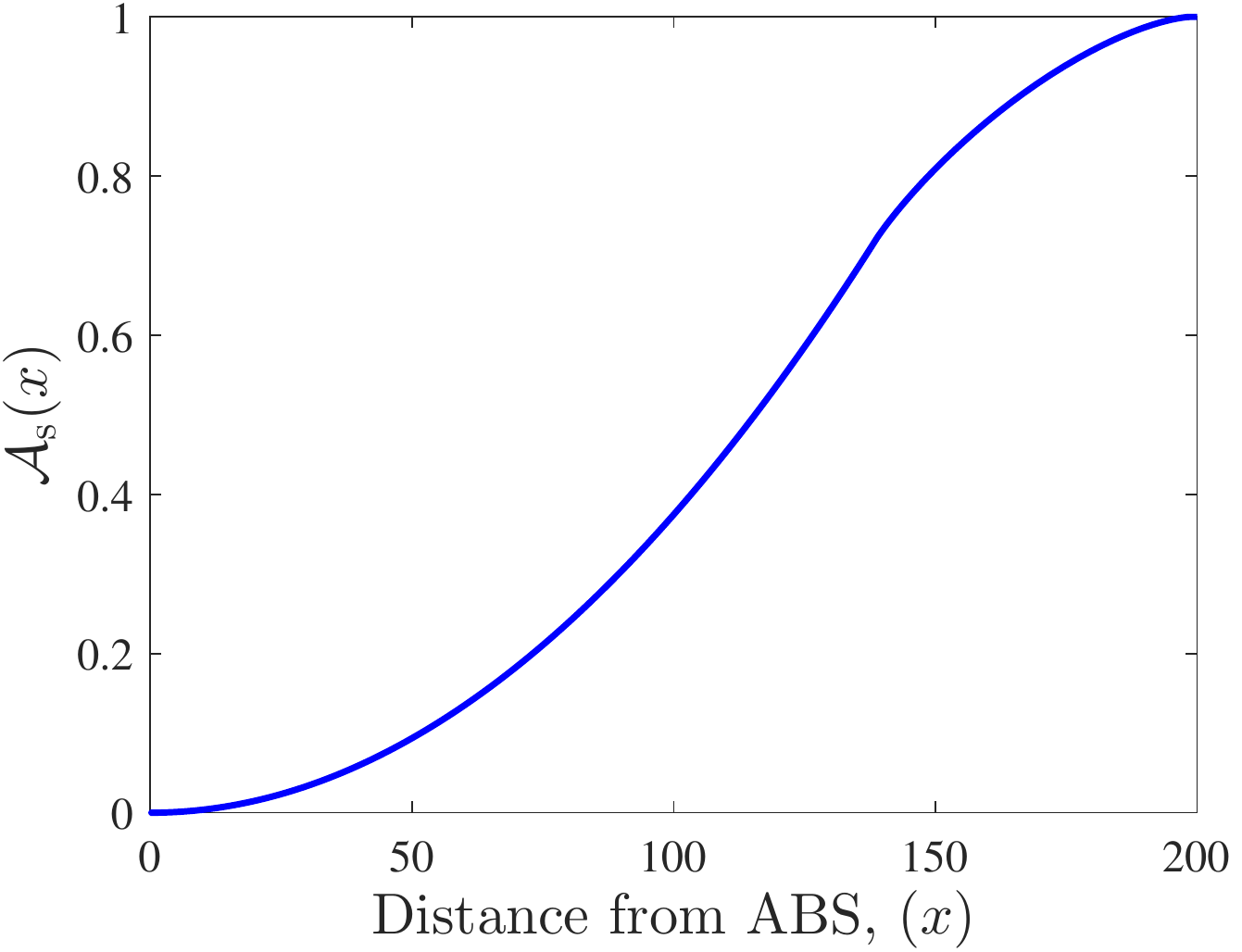}
\caption{Variation of association probability to SBS with distance from ABS.% \chb{to be modified for showing $x'$ and $x''$.}
}\label{fig::sbs::association}
\end{figure}
\subsection{Association Region and Association Probability}
\label{subsec::association::region}
 We first define the association region in the representative user hotspot as follows. Given the representative hotspot is centered at ${\bf x}$, the SBS association region is defined as: ${\cal S}_{\bf x}=\{{\bf x}+{\bf u}\in b({\bf x},R_{\rm s}):P_{\rm m}\|{\bf x}+{\bf u}\|^{-\alpha}<P_{\rm s}u^{-\alpha}\}$ and the ABS association area is  $b({\bf x},R_{\rm s})\cap {\cal S}_{\bf x}^c$. In the following Proposition,  we characterize the shape of ${\cal S}_{\bf x}$. 
\begin{prop}\label{prop::SBS::assocaition::region}
The SBS association region ${\cal S}_{\bf x}$  for the SBS at ${\bf x}$ can be  written as: ${\cal S}_{\bf x} = $
\begin{equation}
\begin{cases}
b\bigg((1-k_p^2)^{-1}{\bf x}, \frac{k_p x}{1-k_p^2}\bigg),&0<x<\frac{k_pR_{\rm s}}{1+k_p},\\
b\bigg((1-k_p^2)^{-1}{\bf x}, \frac{k_p x}{1-k_p^2}\bigg)\cap b({\bf x},R_{\rm s}),&\frac{k_pR_{\rm s}}{1+k_p}\leq x\leq \frac{k_pR_{\rm s}}{1-k_p},\\
b({\bf x},R_{\rm s}), &x>\frac{k_pR_{\rm s}}{1-k_p},
\end{cases}
\label{eq::SBS::assocaition::region}
\end{equation}where $k_p =\bigg(\frac{P_{\rm s}}{P_{\rm m}}\bigg)^{1/\alpha}$.
\end{prop}
\begin{IEEEproof}
Let  ${\bf x} = (X_1, X_2)$ be the Cartesian representation of ${\bf x}$. Let,  ${\cal S}_{\bf x}=\{(X_1+t_1,X_2+t_2)\}$. Then, following the definition of ${\cal S}_{\bf x}$,  $
 P_{\rm m}(t_1^2+t_2^2)^{-\alpha/2}\leq P_{\rm s}((t_1-X_1)^2+(t_2-X_2)^2)^{-\alpha/2}
 \Rightarrow \bigg(t_1 - \frac{X_1}{1-k_p^2} \bigg)^2 +\bigg(t_2 - \frac{X_2}{1-k_p^2} \bigg)^2\leq \bigg(\frac{k_p x}{1-k_p^2}\bigg)^2$.
 Thus, $\{(t_1,t_2)\} = b((1-k_p^2)^{-1}{\bf x}, k_px/(1-k_p^2))$.  Since,  ${\cal S}_{\bf x}$ can not spread beyond $b({\bf x},R_{\rm s})$, 
 ${\cal S}_{\bf x} = b\bigg((1-k_p^2)^{-1}{\bf x}, \frac{k_p x}{1-k_p^2}\bigg)\cap b({\bf x},R_{\rm s})$.
 When $0<x<\frac{k_p}{1+k_p}R_{\rm s}$, $b\bigg((1-k_p^2)^{-1}{\bf x}, \frac{k_p x}{1-k_p^2}\bigg)\subset b({\bf x},R_{\rm s})$. Beyond this limit of $x$, a part of $b((1-k_p^2)^{-1}{\bf x}, k_px/(1-k_p^2))$ lies outside of $b({\bf x},R_{\rm s})$.   Finally, when $x>\frac{k_p}{1-k_p}R_{\rm s}$, $b((1-k_p^2)^{-1}{\bf x}, k_px/(1-k_p^2))\supset b({\bf x},R_{\rm s})$.
\end{IEEEproof}
 This formulation of ${\cal S}_{\bf x}$ is illustrated in Fig.~\ref{fig::association::construction}. 
We now compute  the  SBS association probability as follows. 
\begin{lemma}\label{lemm::association::sbs}
 Conditioned on the fact that the user belongs to the  hotspot  at ${\bf x}$, the association probability to SBS is given by: ${\cal A}_{\rm s}({\bf x}) ={\cal A}_{\rm s}(x)=$
%\begin{multline}
\begin{align}%\label{eq::association::sbs}
& \int_0^{2\pi}\frac{\bigg(\min\big(R_{\rm s},x\frac{k_p(\sqrt{1-k_p^2.\sin^2\xi}+k_p\cos\xi)}{1-k_p^2}\big)\bigg)^2}{2\pi R_{\rm s}^2}{\rm d}\xi\label{eq::association::abs}\\
&= \begin{cases} \frac{k_p^2x^2}{(1-k_p^2)^2R_{\rm s}^2}&\text{if }0<x< \frac{k_p}{1+k_p}R_{\rm s},\\
\frac{{\cal C}\big(R_{\rm s},\frac{k_p x}{1-k_p^2}, \frac{k_p^2 x}{1-k_p^2}\big)}{\pi R_{\rm s}^2}&\text{if }\frac{k_p}{1+k_p}R_{\rm s}\leq x\leq \frac{k_p}{1-k_p}R_{\rm s},\\
1&\text{if }x>\frac{k_p}{1-k_p}R_{\rm s}\label{eq::association::abs::2}
\end{cases},
\end{align}
where \begin{multline*}{\cal C}(r_1,r_2,d)= r_1^2\tan^{-1}\bigg(\frac{t}{d^2+r_1^2-r_2^2}\bigg)\\+r_2^2\tan^{-1}\bigg(\frac{t}{d^2-r_1^2+r_2^2}\bigg)-{\frac{t}{2}}
\end{multline*} is the area of intersection of two intersecting circles of radii $r_1$, and $r_2$ and distance between centers $d$ with $ t=(d+r_1+r_2)^{\frac{1}{2}}(d+r_1-r_2)^{\frac{1}{2}}(d-r_1+r_2)^{\frac{1}{2}}(-d+r_1+r_2)^{\frac{1}{2}}$. The association probability to the ABS is given by ${\cal A}_{\rm m}({ x})= 1-{\cal A}_{\rm s}({x})$.
\end{lemma}
\begin{IEEEproof}
Conditioned on the location of the hotspot center at ${\bf x}$, ${\cal A}_{\rm s}({\bf x})= \nbbP({\cal E} = 1|{\bf x})=$
\begin{align*}
&\nbbE[{\bf 1}(P_{\rm m}\|{\bf x}+{\bf u}\|^{-\alpha}<P_{\rm s}\|{\bf u}\|^{-\alpha})|{\bf x}]=\nbbP({\bf x}+{\bf u}\in {\cal S}_{\bf x}|{\bf x})\\
&=\nbbP(P_{\rm m}(x^2+u^2+2xu\cos\xi)^{-\alpha/2}<P_{\rm s}u^{-\alpha})|x)\\&=
\nbbP(u^2(1-k_p^2)-2x\cos\xi k_p^2u-k_p^2x^2<0|x)\\&\myeq{a}\nbbP\bigg(u\in\bigg(0, \frac{xk_p\big(\sqrt{1-k_p^2\sin^2\xi}+k_p\cos\xi\big)}{1-k_p^2}\bigg), \\&\qquad\qquad\xi\in(0, 2\pi]\bigg|x\bigg)\\&
=\int_{0}^{2\pi}\int_{0}^{R_{\rm s}}{\bf 1}\bigg(0\leq u < \frac{xk_p\big(\sqrt{1-k_p^2\sin^2\xi}+k_p\cos\xi\big)}{1-k_p^2}\bigg)\\&\times f_U(u){\rm d}u \frac{1}{2\pi}{\rm d}{\xi},
%\myeq{b}\nbbE\bigg[\frac{\min\bigg(R_{\rm s},\frac{xk_p\sqrt{1-k_p^2\sin^2\xi+k_p\cos\xi}}{1-k_p^2}\bigg)}{R_{\rm s}^2}\big| x\bigg], %\label{eq::associtation::prob::proof::intermediate}
\end{align*}
where $\xi = \arg({\bf u}-{\bf x})$ and is uniformly distributed in $(0,2\pi]$. Here, (a) follows from solving the quadratic inequality inside the indicator function. The last step   follows from deconditioning over ${u}$ and $\xi$. Finally, \eqref{eq::association::abs}   is obtained  by {evaluating} the integration over $u$. Note that, due to angular symmetry, ${\cal A}_{\rm s}({\bf x})={\cal A}_{\rm s}({ x})$. 
Alternatively, $${\cal A}_{\rm s}(x)=\int_{{\cal S}_{\bf x}}f_{U}(u){\rm d}u\frac{1}{2\pi}{\rm d}{\xi}=\frac{|{\cal S}_{\bf x}|}{\pi R_{\rm s}^2}.$$ The final result in \eqref{eq::association::abs::2} {is} obtained  by using Proposition~\ref{prop::SBS::assocaition::region}.
\end{IEEEproof}
In 
 \figref{fig::sbs::association}, we plot ${\cal A}_{\rm s}(x)$ as a function of $x$. 
We now evaluate the coverage probability of a typical user which is the probability of the occurrence of the events defined in \eqref{eq::coverage::def}.  
\begin{theorem}%[Coverage probability]
\label{thm::coverage::probability}The coverage probability
 is given by:
 \begin{equation}\label{eq::coverage::probability}
 \pc =\int\limits_{0}^{R-R_{\rm s}}\big(\pc_{\rm s}(\theta_1,\theta_2|x)+\pc_{\rm m}(\theta_3|x)\big)f_X(x){\rm d}x,
 \end{equation}
 where $\pc_{\rm s} (\theta_1,\theta_2|x) =$
 \begin{multline*}
  \int\limits_{0}^{2\pi}\int\limits_{0}^{u_{\max}(x,\xi)} 
\bigg(p({x}) 
F_h\bigg(\frac{ x^{\alpha_L}\beta{\tt N}_0W\theta_1}{P_{\rm m}G^2 },m_L\bigg) +(1-p({x}))\\
\times F_h\bigg(\frac{ x^{\alpha_{NL}}\beta{\tt N}_0W\theta_1}{P_{\rm m}G^2 },m_{NL}\bigg)\bigg)\bigg(p({u})
F_h\bigg(\frac{ u^{\alpha_L}\beta{\tt N}_0W\theta_2}{P_{\rm s}G },m_{L}\bigg) \\+(1-p({u})) F_h\bigg(\frac{ u^{\alpha_{NL}}\beta{\tt N}_0W\theta_2}{P_{\rm s} G},m_{NL}\bigg)\bigg)\frac{f_{ U}(u)}{2\pi}{\rm d}{u}\:{\rm d}{\xi},
\end{multline*}
where 
  $u_{\max}(x,\xi) =  \min\bigg(R_{\rm s}, x k_p\frac{\sqrt{(1-k_p^2\sin^2\xi)}+k_p\cos\xi)}{1-k_p^2}\bigg)$ and $F_{h}(\cdot)$ is the {complementary cumulative distribution function (CCDF) of Gamma distribution}, %, expressed as:
%\begin{equation}\label{eq::ccdf::gamma}
%F_{h}(z,\beta) = 1-\frac{1}{\Gamma(\beta)}\int\limits_{0}^{z/\beta}t^{\beta-1}e^{-t}{\rm d}t,
%\end{equation}  
    and $\pc_{\rm m} (\theta_3|x)=$
\begin{multline*}
 \int\limits_{0}^{2\pi}\int\limits_{u_{\max}(x,\xi)}^{R_{\rm s}} {\bigg( p(\kappa(x,u,\xi)) 
F_h\bigg(\frac{{\kappa(x,u,\xi)}^{\alpha_L}\beta{\tt N}_0W\theta_3}{P_{\rm m}G },m_L\bigg)}\\ +
(1-p(\kappa(x,u,\xi))) F_h\bigg(\frac{ {\kappa(x,u,\xi)}^{\alpha_{NL}}\beta{\tt N}_0W\theta_3}{P_{\rm m}G },m_{NL}\bigg)\bigg)\\\times\frac{f_{ U}(u){\rm d}u\:{\rm d}\xi}{2\pi},
\end{multline*}
\end{theorem}
where $\kappa(x,u,\xi)=({x^2+u^2+2 x u \cos\xi})^{1/2}$.
\begin{IEEEproof}
%\begin{figure*}
%
%\begin{multline}
%\pc_{\rm m} (\theta_3|x)= \int\limits_{0}^{2\pi}\int\limits_{u_{\max}(x,\xi)}^{R_{\rm s}} \bigg[ p(({x^2+u^2+2 x u \cos\xi})^{1/2}) 
%F_h\bigg(\frac{{(x^2+u^2+2 x u \cos\xi)}^{\alpha_L/2}\sigma_N^2\theta_3}{P_{\rm s}G },m_L\bigg)\\ +
%(1-p({(x^2+u^2+2 x u \cos\xi)}^{1/2})) F_h\bigg(\frac{ {(x^2+u^2+2 x u \cos\xi)}^{\alpha_{NL}/2}\sigma_N^2\theta_3}{P_{\rm s}G },m_{NL}\bigg){\rm d}u\:{\rm d}\xi\bigg].
%\end{multline}
%\end{figure*}
See Appendix~\ref{app::coverage::probability}.
\end{IEEEproof}
As expected, coverage probability  is the summation of two terms, each corresponding to the probability of occurrences of the two mutually exclusive events appearing in \eqref{eq::coverage::def}. 

%The expression of coverage probability serves as the first fundamental step towards the derivation of rate-coverage probability.
\subsection{Load distributions}\label{subsec::load::dist}
While the load distributions for the PPP-based models are well-understood~\cite{OffloadingSingh,SinghKulkarniSelfBackhaul}, they are not directly applicable to the 3GPP-inspired finite model used in this paper. Consequently, in this Section, we provide a novel approach to characterize the ABS and SBS load for this model. 
  As we saw in \eqref{eq::rate_abs_access}, the load on the ABS has  two components, one is due to the contribution of the number of users of the {representative} hotspot connecting to the 
ABS (denoted by $N_{{\bf x}}^{\rm ABS}$)  and the other is due to the macro users of  {the} other clusters, which we lump into a single random variable, $N_{\rm o}^{\rm ABS} = \sum_{i=1}^{n-1}N_{{\bf x}_i}^{\rm ABS}$. On the other hand, $N_{\bf x}^{\rm SBS}$ and $N_{\rm o}^{\rm SBS}=\sum_{i=1}^{n-1}N_{{\bf x}_i}^{\rm SBS}$ respectively denote the load on the SBS at ${\bf x}$ and sum load of all SBSs except the one at ${\bf x}$. 
 First, we obtain the PMFs of $N_{{\bf x}}^{\rm ABS}$ and $N_{{\bf x}}^{\rm SBS}$ using the fact that given the location of the representative hotspot centered at ${\bf x}$, {each user  belongings to  the association regions ${\cal S}_{\bf x}$ or $b({\bf x},R_{\rm s})\cap {{\cal S}}_{\bf x}^c$ according to an i.i.d. Bernoulli random variable.} 
 
 %Before we present these PMFs, we redefine the notion of typical user and representative cluster  for \case~$2$ following the consequence of length-biased sampling from a population of Poisson distributed individuals~\cite{chiu2013stochastic} in the following Remark. 
% \begin{remark}\label{rem::typicality::case2} Recall that we defined the typical user as a user  sampled uniformly at random from the population.   
 %For the interest of tractability of  the distribution of $N_{\bf x}$ in \case~$2$, we first sample the representative hotspot centered at ${\bf x}$ uniformly at random from $n$ hotspots and then add  the typical user at ${\bf x}+{\bf u}$, where $\bf u$ follows the PDF in \eqref{eq::users::distribution}. Changing the notion of typicality for \case~$2$ has negligible impact on $N_{\bf x}$ for a moderately large population size ($\sum_{i-1}^nN_{{\bf x}_i}$) and the two notions of typicality are exactly equivalent when  $\sum_{i-1}^nN_{{\bf x}_i}\to\infty$.    
% \end{remark} 
 %Following Remark~\ref{rem::typicality::case2}, $N_{\bf x}$ follows a weighted Poisson distribution~\cite{chiu2013stochastic}: $
%\nbbP(N_{\bf x} = k ) = \frac{\bar{m}^{k-1}e^{-\bar{m}}}{(k-1)!}$, where,  $k\in\nbbZ^+$. 
\begin{lemma}\label{lemm::load::characterization::abs}
Given the fact that the representative  hotspot is centered at ${\bf x}$, load on the ABS  due to the macro users in the hotspot at ${\bf x}$ ($N^{\rm ABS}_{{\bf x}}$) and load on the SBS at $\bf x$  ($N^{\rm SBS}_{\bf x}$) are distributed as follows:

\case~$1$ ($N_{{\bf x}_i} = \bar{m}$, $\forall\ i=1,\dots,n$).
\begin{subequations}
\begin{alignat}{2}
&\nbbP(N^{\rm ABS}_{{\bf x}}=k|{\bf x})= {\bar{m}-1\choose k-1}{\cal A}_{\rm m}(x)^{k-1}{\cal A}_{\rm s}(x)^{\bar{m}-k}\label{eq::load::abs::load_x::fixedN},\\
&\nbbP(N^{\rm SBS}_{{\bf x}}=k|{\bf x})= {\bar{m}-1\choose k-1}{\cal A}_{\rm s}(x)^{k-1}{\cal A}_{\rm m}(x)^{\bar{m}-k},\label{eq::load::sbs::load_x::fixedN}
\end{alignat}
\end{subequations}
where $k=1,2,\dots,\bar{m}$.

\case~$2$ ($N_{{\bf x}_i} \stackrel{i.i.d.}{\sim} {\tt Poisson}(\bar{m})$, $\forall\ i=1,\dots,n$).
\begin{subequations}
\begin{alignat}{2}
&\nbbP(N^{\rm ABS}_{{\bf x}}=k|{\bf x})= \frac{(\bar{m}{\cal A}_{\rm m}(x))^{k-1}}{(k-1)!}e^{-\bar{m}{\cal A}_{\rm m}(x)}\label{eq::load::abs::load_x::PoissonN},\\
&\nbbP(N^{\rm SBS}_{{\bf x}}=k|{\bf x})= \frac{(\bar{m}{\cal A}_{\rm s}(x))^{k-1}}{(k-1)!}e^{-\bar{m}{\cal A}_{\rm s}(x)},\label{eq::load::sbs::load_x::PoissonN}
\end{alignat}
\end{subequations}
where $k\in\nbbZ^{+}$.
\end{lemma}
\begin{IEEEproof}See Appendix~\ref{app::load::characterization::abs}.
\end{IEEEproof}
{We present the first moments of these two load variables in the following Corollary which will be required for the evaluation of the rate coverage for the average load-based partition and the derivation of easy-to-compute approximations  of rate coverage in the sequel.} 
\begin{cor}\label{cor::mean::representative::loads}
{The conditional means of $N_{\bf x}^{\rm ABS}$ and $N_{\bf x}^{\rm SBS}$ given the center of the representative hotspot at $\bf x$ are}
\begin{align*}
&\text{\case~$1$: }\nbbE[N_{\bf x}^{\rm ABS}] = (\bar{m}-1){\cal A}_{\rm m}(x)+1, \nbbE[N_{\bf x}^{\rm SBS}] = \\&\qquad\qquad\qquad\qquad(\bar{m}-1){\cal A}_{\rm s}(x)+1,\\
&\text{\case~$2$: }\nbbE[N_{\bf x}^{\rm ABS}] = \bar{m}{\cal A}_{\rm m}(x)+1, \nbbE[N_{\bf x}^{\rm SBS}] = \bar{m}{\cal A}_{\rm s}(x)+1.
\end{align*}
\end{cor}
We now obtain the PMFs of $N_{\rm o}^{\rm ABS}$ and $N_{\rm o}^{\rm SBS}$  in the following Lemma. {Note that, since ${\bf x}_i$-s are i.i.d., $N_{\rm o}^{\rm ABS}$ and $N_{\rm o}^{\rm SBS}$ are independent of ${\bf x}$.}   In what follows, the exact PMF of $N_{{\rm o}}^{\rm ABS}$ ($N_{{\rm o}}^{\rm SBS}$) is in the form of  $(n-1)$-fold discrete convolution and hence is not computationally efficient beyond very small values of $n$.  We present an {alternate} easy-to-use expression of this PMF by invoking central limit theorem (CLT). In the numerical {results} Section, we verify that this approximation is {tight} even for moderate   values of $n$.
\begin{lemma}\label{lemm::load::characterization::others}
Given the fact that the typical user belongs to a hotspot at ${\bf x}$, load on the ABS due to all other $n-1$ hotspots   is distributed as:
{$
 \frac{N_{\rm o}^{\rm ABS}-\upsilon_{\rm m}}{\sigma_{\rm m}}\sim{\cal N}(0,1)\text{ (for large $n$)}$}
 and sum of the loads on the other SBSs at  ${\bf x}_1,{\bf x}_2,\dots,{\bf x}_{n-1}$  is distributed as:
$ \frac{N_{\rm o}^{\rm SBS}-\upsilon_{\rm s}}{\sigma_{\rm s}}\sim{\cal N}(0,1) \text{ (for large $n$)}$, where ${\cal N}(0,1)$ denotes the standard normal distribution,
$
\upsilon_{\rm m}=(n-1)\bar{m}\nbbE[{\cal A}_{\rm m}(X)], \upsilon_{\rm s}=(n-1)\bar{m}\nbbE[{\cal A}_{\rm s}(X)],%\label{eq::load::others::fixedN::mean}
$ 
and  
\begin{align*}
&\text{for \case~$1$, }\sigma_{\rm m}^2=(n-1)\big[\bar{m}\nbbE[{\cal A}_{\rm m}(X){\cal A}_{\rm s}(X)]\\&\qquad\qquad\qquad\quad+\bar{m}^2{\rm Var}[{\cal A}_{\rm m}(X)]\big]=\sigma_{\rm s}^2, \\
&\text{for \case~$2$, }\sigma_{\rm m}^2=(n-1)\big[\bar{m}\nbbE[{\cal A}_{\rm m}(X)]+\bar{m}^2{\rm Var}[{\cal A}_{\rm m}(X)]\big],\\
&\sigma_{\rm s}^2=(n-1)\big[\bar{m}\nbbE[{\cal A}_{\rm s}(X)]+\bar{m}^2{\rm Var}[{\cal A}_{\rm s}(X)]\big].
\end{align*}
Here,
\begin{multline*}%\label{eq::mean_variance::association}
\nbbE[{\cal A}_{\rm m}(X)]=\int_{0}^{R-R_{\rm s}}{\cal A}_{\rm m}(x)f_{X}(x){\rm d}x,\text{ and } \\
{\rm Var}[{\cal A}_{\rm m}(X)] = \int\limits_{0}^{R-R_{\rm s}}\big({\cal A}_{\rm m}(x)\big)^2f_X(x){\rm d}x - (\nbbE[{\cal A}_{\rm m}(X)])^2,
\end{multline*}
and $\nbbE[{\cal A}_{\rm s}(X)]$, ${\rm Var}[{\cal A}_{\rm s}(X)]$ can be similarly obtained by replacing ${\cal A}_{\rm m}(X)$ with ${\cal A}_{\rm s}(X)$ in the above expressions. 
\end{lemma}
\begin{IEEEproof}
See Appendix~\ref{app::load::characterization::others}.
\end{IEEEproof}
\subsection{Rate Coverage Probability}
 We first define the downlink rate coverage probability (or simply, rate coverage) as follows. 
\begin{ndef}[Rate coverage probability]\label{def::rate::coverage} The rate coverage probability of a link with BW $\tilde{W}$ is defined as the probability that the maximum achievable data rate (${\cal R}$) exceeds a certain threshold $\rho$, i.e., $\nbbP({\cal R}>\rho)  =$
\begin{align}
 \nbbP\bigg(\tilde{W}\log_{2}(1+\snr)>\rho\bigg) = \nbbP(\snr>2^{{\rho}/{\tilde{W}}}-1).
\end{align}
\end{ndef}
Hence, we see that the rate coverage probability is the coverage probability evaluated at a modified $\snr$-threshold. We now evaluate the rate coverage probability  for different backhaul BW partition strategies for {a general distribution of $N_{{\bf x}_i}$ and $N_{{\bf x}}$ in the following Theorem.  We later specialize this result for \case s~$1$ and $2$ for numerical evaluation.}
\begin{theorem}\label{thm::rate::cov::equal::partition}The rate coverage probability for a target data rate $\rho$  is given by:
\begin{equation}
\pr = \pr_{\rm m} + \pr_{\rm s},
\end{equation}
where $\pr_{\rm m}$ ($\pr_{\rm s}$) denotes the ABS rate coverage (SBS rate coverage) which is the probability that the typical user is receiving data rate greater than or equal to $\rho$ and is served by the ABS (SBS). The ABS rate coverage is given by:
\begin{multline}\label{eq::rate::cov::macro}
\pr_{\rm m}=\int\limits_{-\infty}^{\infty}\int\limits_{0}^{R-R_{\rm s}}\nbbE_{N_{\bf x}^{\rm ABS}}\bigg[\pc_{\rm m}\bigg(2^{\frac{\rho(t+N_{\bf x}^{\rm ABS})}{W_{\rm a}}}-1|x\bigg)\bigg]\\\times
f_X(x){\rm d}x \frac{1}{\sigma_{\rm m}\sqrt{2\pi}}e^{-\frac{(t-\upsilon_{\rm m})^2}{2\sigma_{\rm m}^2}}{\rm d}t.
\end{multline}
The SBS rate coverage depends on the backhaul BW partition strategy. For equal partition, 
\begin{equation}
\pr_{\rm s} =\int\limits_{0}^{R-R_{\rm s}} \nbbE_{N_{\bf x}^{\rm SBS}}\bigg[\pc_{\rm s}\bigg(2^{\frac{\rho n N^{\rm SBS}_{\bf x}}{W_{\rm b}}}-1,2^{\frac{\rho N^{\rm SBS}_{\bf x}}{W_{\rm a}}}-1\big|x\bigg)\bigg]f_X(x){\rm d}x,\label{eq::rate::cov::sbs::eq::partition}
\end{equation}
\vspace{-1em} 
 for instantaneous load-based partition, 
 \begin{multline}
\label{eq::rate::cov::inst-load::sbs}
\pr_{\rm s}= \int\limits_{-\infty}^{\infty}\int\limits_{0}^{R-R_{\rm s}}\nbbE_{N_{\bf x}^{\rm SBS}}\bigg[\pc_{\rm s}\bigg(2^{\frac{\rho( N^{\rm SBS}_{\bf x}+t)}{W_{\rm b}}}-1,2^{\frac{\rho  N^{\rm SBS}_{\bf x}}{W_{\rm a}}}-1\big| x\bigg)\bigg]\\\times f_{X}(x){\rm d}x \frac{1}{\sigma_{\rm s}\sqrt{2\pi}}e^{-\frac{(t-\upsilon_{\rm s})^2}{2\sigma_{\rm s}^2}}{\rm d}t,
\end{multline}
and for average load-based partition,
\begin{multline}
\label{eq::rate::cov::avg-load::sbs}
\pr_{\rm s}= \int\limits_{-\infty}^{\infty}\int\limits_{0}^{R-R_{\rm s}}\nbbE_{N_{\bf x}^{\rm SBS}}\bigg[\pc_{\rm s}\bigg(2^{\frac{\rho N^{\rm SBS}_{\bf x}\left(\nbbE[N_{\bf x}^{\rm SBS}]+\bar{m}t\right)}{W_{\rm b}\nbbE[N_{\bf x}^{\rm SBS}]}}-1,\\2^{\frac{\rho  N^{\rm SBS}_{\bf x}}{W_{\rm a}}}-1\big| x\bigg)\bigg]f_{X}(x){\rm d}x \frac{e^{-\frac{\big(t-(n-1)\nbbE[{\cal A}_{\rm s}(X)]\big)^2}{2(n-1){\rm Var}[{\cal A}_{\rm s}(X)]}}}{\sqrt{2\pi(n-1){\rm Var}[{\cal A}_{\rm s}(X)]}}{\rm d}t.
\end{multline}
%and $\pr_{\rm a}$ is given by \eqref{eq::rate::cov::macro::fixedN}.
\end{theorem}
\begin{IEEEproof}See Appendix~\ref{app::rate::cov::equal::partition}.
\end{IEEEproof}
{Note that the key enabler of the expression of $\pr$ in Theorem~\ref{thm::rate::cov::equal::partition} is the fact that the system is considered to be noise-limited. Including the SBS interference into analysis is not straightforward from this point since it would involve coupling between the coverage probability and load since both are dependent on the locations of the other $n-1$ SBSs.}      
Having derived the exact expressions of rate coverage in Theorem~\ref{thm::rate::cov::equal::partition}, we present   approximations of these expressions by replacing (i) $N_{\bf x}^{\rm ABS}$ in $\pr_{\rm m}$ with its mean $\nbbE[N_{\bf x}^{\rm ABS}]$, and (ii) $N_{\bf x}^{\rm SBS}$ in $\pr_{\rm s}$ with its mean $\nbbE[N_{\bf x}^{\rm SBS}]$ in the following Lemma.  
\begin{lemma}\label{lemm::approximation} {The ABS  rate coverage can be approximated as} 
\begin{multline}
\label{eq::rate::cov::macro::approx::fixedN}
\pr_{\rm m}=\int\limits_{-\infty}^{\infty}\int\limits_{0}^{R-R_{\rm s}}\pc_{\rm m}\bigg(2^{\frac{\rho(t+\nbbE[N_{\bf x}^{\rm ABS}])}{W_{\rm a}}}-1|x\bigg)f_X(x){\rm d}x \\ 
\times \frac{e^{-\frac{(t-\upsilon_{\rm m})^2}{2\sigma_{\rm m}^2}}}{\sigma_{\rm m}\sqrt{2\pi}}{\rm d}t.
\end{multline}
The SBS rate coverage can be approximated as follows. For  equal partition, 
\begin{align}
\pr_{\rm s} = \int\limits_{0}^{R-R_{\rm s}}\pc_{\rm s}\bigg(2^{\frac{\rho n\nbbE[ N^{\rm SBS}_{\bf x}]}{W_{\rm b}}}-1,2^{\frac{\rho \nbbE[N^{\rm SBS}_{\bf x}]}{W_{\rm a}}}-1\big|x\bigg)f_X(x){\rm d}x,\label{eq::rate::cov::sbs::eq::partition::approx}
\end{align}
 for instantaneous load-based partition, 
\begin{multline}%\label{eq::rate::cov::inst-load::sbs::approx}
\label{eq::rate::cov::inst-load::sbs::approx}
\pr_{\rm s}= \int\limits_{-\infty}^{\infty}\int\limits_{0}^{R-R_{\rm s}}\pc_{\rm s}\bigg(2^{\frac{\rho( \nbbE[N^{\rm SBS}_{\bf x}]+t)}{W_{\rm b}}}-1,2^{\frac{\rho \nbbE[  N^{\rm SBS}_{\bf x}]}{W_{\rm a}}}-1\big| x\bigg)\\\times f_{X}(x){\rm d}x  \frac{e^{-\frac{(t-\upsilon_{\rm s})^2}{2\sigma_{\rm s}^2}}}{\sigma_{\rm s}\sqrt{2\pi}}{\rm d}t,
\end{multline}
and for average load-based partition,
\begin{multline}%\label{eq::rate::cov::inst-load::sbs::approx}
\label{eq::rate::cov::avg-load::sbs::approx}
\pr_{\rm s}=\int\limits_{-\infty}^{\infty}\int\limits_{0}^{R-R_{\rm s}}\pc_{\rm s}\bigg(2^{\frac{\rho \left(\nbbE[N_{\bf x}^{\rm SBS}]+\bar{m}t\right)}{W_{\rm b}}}-1,2^{\frac{\rho  \nbbE[ N^{\rm SBS}_{\bf x}]}{W_{\rm a}}}-1\big| x\bigg)\\\times f_{X}(x){\rm d}x \frac{e^{-\frac{\big(t-(n-1)\nbbE[{\cal A}_{\rm s}(X)]\big)^2}{2(n-1){\rm Var}[{\cal A}_{\rm s}(X)]}}}{\sqrt{2\pi(n-1){\rm Var}[{\cal A}_{\rm s}(X)]}}{\rm d}t.
\end{multline}
%and $\pr_{\rm a}$ is given by \eqref{eq::rate::cov::macro::fixedN}.
\end{lemma}
\begin{table*}[t]
\centering
\caption{{Key system parameters}}%\vspace{-0.5cm}
\label{tab::parameters}%\scalebox{0.8}
{
\begin{tabular}{|l|l|l|}
\hline
Notation & Parameter & Value \\ \hline
%$W$ & mm-wave bandwidth & 2 GHz \\ \hline
$P_{\rm m},\ P_{\rm s}$ & BS transmit powers  & 50, 20 dBm \\ \hline
$\alpha_L, \alpha_{NL}$ & Path-loss exponent &  2.0,  3.3\\ \hline
$\beta$ & Path loss at 1 m & 70 dB \\\hline
$G$ & Main lobe gain & 18 dB \\ \hline
$\mu$ & LOS range constant & 170 m \\ \hline
${\tt N}_0W$ & Noise power & \begin{tabular}[c]{@{}l@{}}-174 dBm/Hz+ $10\log_{10} W$ \\+10 dB {(noise-figure)}\end{tabular} \\ \hline
$m_L,m_{NL}$& Parameter of Nakagami distribution& 2,  3\\\hline
$R$, $R_{\rm s}$& Macrocell and hotspot radius & 200 m, 30 m\\
\hline
$\bar{m}$& Average number of users per hotspot & 5\\\hline
$\rho$& Rate threshold & 50 Mbps\\
\hline
\end{tabular}
}
\end{table*}
{We now specialize the result of Theorem~\ref{thm::rate::cov::equal::partition} for  \case s~$1$ and $2$ in the following Corollaries.}  
\begin{cor}\label{cor::rate::cov::fixedN}For \case~$1$, i.e., when $N_{{\bf x}_i} = \bar{m}$, $\forall\ i=1,\dots,n$, the ABS rate coverage is 
\begin{multline}\label{eq::rate::cov::macro::fixedN}
\pr_{\rm m} = \sum\limits_{k=1}^{\bar{m}}{{\bar{m}-1} \choose k-1}\int\limits_{-\infty}^{\infty}\int\limits_{0}^{R-R_{\rm s}}\pc_{\rm m}\bigg(2^{\frac{\rho(t+k)}{W_{\rm a}}}-1|x\bigg)\\\times{\cal A}_{\rm m}(x)^{k-1}{\cal A}_{\rm s}(x)^{\bar{m}-k} f_X(x){\rm d}x \frac{1}{\sigma_{\rm m}\sqrt{2\pi}}e^{-\frac{(t-\upsilon_{\rm m})^2}{2\sigma_{\rm m}^2}}{\rm d}t.
\end{multline}The SBS rate coverages for the three backhaul BW parition strategies are expressed as follows. (i) For equal partition, 
\begin{multline}
\pr_{\rm s} = 
\sum\limits_{k = 1}^{\bar{m}}{{\bar{m}-1} \choose k-1}\int\limits_{0}^{R-R_{\rm s}}\pc_{\rm s}\bigg(2^{\frac{\rho n k}{W_{\rm b}}}-1,2^{\frac{\rho k}{W_{\rm a}}}-1\big|x\bigg)\\\times  {\cal A}_{\rm s}(x)^{k-1}{\cal A}_{\rm m}(x)^{\bar{m}-k}f_X(x){\rm d}x,\label{eq::rate::cov::sbs::fixedN::eq::partition}
\end{multline}
(ii) for instantaneous load-based partition, 
\begin{multline}\label{eq::rate::cov::inst-load::sbs::fixedN}\pr_{\rm s} = \sum\limits_{k = 1}^{\bar{m}}{{\bar{m}-1} \choose k-1}\int\limits_{-\infty}^{\infty}\int\limits_{0}^{R-R_{\rm s}}\pc_{\rm s}\bigg(2^{\frac{\rho(k+t)}{W_{\rm b}}}-1,2^{\frac{\rho k}{W_{\rm a}}}-1\big| x\bigg)\\\times{\cal A}_{\rm s}(x)^{k-1}{\cal A}_{\rm m}(x)^{\bar{m}-k}f_{X}(x){\rm d}x \frac{e^{-\frac{(t-\upsilon_{\rm s})^2}{2\sigma_{\rm s}^2}}}{\sigma_{\rm s}\sqrt{2\pi}}{\rm d}t,
\end{multline}
and (iii) for average load-based partition,
\begin{multline}\label{eq::rate::cov::avg-load::sbs::fixedN}
\pr_{\rm s}=\sum\limits_{k = 1}^{\bar{m}}{{\bar{m}-1} \choose k-1}\int\limits_{-\infty}^{\infty}\int\limits_{0}^{R-R_{\rm s}}\pc_{\rm s}\bigg(2^{\frac{\rho k\left(1+(\bar{m}-1){\cal A}_{\rm s}(x)+\bar{m} t\right)}{W_{\rm b}(1+(\bar{m}-1){\cal A}_{\rm s}(x))}}-1,\\2^{\frac{\rho  k}{W_{\rm a}}}-1\big| x\bigg){\cal A}_{\rm s}(x)^{k-1}{\cal A}_{\rm m}(x)^{\bar{m}-k}\times f_{X}(x){\rm d}x\\\times  \frac{e^{-\frac{\big(t-(n-1)\nbbE[{\cal A}_{\rm s}(X)]\big)^2}{2(n-1){\rm Var}[{\cal A}_{\rm s}(X)]}}}{\sqrt{2\pi(n-1){\rm Var}[{\cal A}_{\rm s}(X)]}}{\rm d}t.
\end{multline}
%and $\pr_{\rm a}$ is given by \eqref{eq::rate::cov::macro::fixedN}.
\end{cor}
\begin{IEEEproof}The result can be obtained from Theorem~\ref{thm::rate::cov::equal::partition} by using the PMFs of $N_{\bf x}^{\rm ABS}$, $N_{\bf x}^{\rm SBS}$, $N_{o}^{\rm ABS}$ and $N_{o}^{\rm ABS}$ from Lemmas~\ref{lemm::load::characterization::abs} and \ref{lemm::load::characterization::others} and substituting $\nbbE[N_{\bf x}^{\rm SBS}]$ {from} Corollary~\ref{cor::mean::representative::loads} for \case~$1$.
\end{IEEEproof}
\begin{cor}\label{cor::rate::cov::PoissonN}For \case~$2$, i.e., when $N_{{\bf x}_i} \stackrel{i.i.d.}{\sim} {\tt Poisson}(\bar{m})$, $\forall\ i=1,\dots,n$, the ABS rate coverage is expressed  as
\begin{multline}\label{eq::rate::cov::macro::PoissonN}
\pr_{\rm m} = \sum\limits_{k=1}^{\infty}\frac{\bar{m}^{k-1}}{(k-1)!}
\int\limits_{-\infty}^{\infty}\int\limits_{0}^{R-R_{\rm s}}
({\cal A}_{\rm m}(x))^{k-1}e^{-\bar{m}{\cal A}_{\rm m}(x)}\\\times 
\pc_{\rm m}
\bigg(2^{\frac{\rho(t+k)}{W_{\rm a}}}-1|x\bigg) f_X(x){\rm d}x \frac{1}{\sigma_{\rm m}\sqrt{2\pi}}e^{-\frac{(t-\upsilon_{\rm m})^2}{2\sigma_{\rm m}^2}}{\rm d}t.
\end{multline}The SBS rate coverages for the three backhaul BW parition strategies are expressed as follows. (i) For equal partition, 
\begin{multline}
\pr_{\rm s} = 
\sum\limits_{k=1}^{\infty}\frac{\bar{m}^{k-1}}{(k-1)!}
\int\limits_{0}^{R-R_{\rm s}}
({\cal A}_{\rm s}(x))^{k-1}e^{-\bar{m}{\cal A}_{\rm s}(x)}\\\times 
\pc_{\rm s}\bigg(2^{\frac{\rho n k}{W_{\rm b}}}-1,2^{\frac{\rho k}{W_{\rm a}}}-1\big|x\bigg)f_X(x){\rm d}x,\label{eq::rate::cov::sbs::PoissonN::eq::partition}
\end{multline}
(ii) for instantaneous load-based partition, 
\begin{multline}\label{eq::rate::cov::inst-load::sbs::PoissonN}\pr_{\rm s} = \sum\limits_{k=1}^{\infty}\frac{\bar{m}^{k-1}}{(k-1)!}
\int\limits_{-\infty}^{\infty}\int\limits_{0}^{R-R_{\rm s}}
({\cal A}_{\rm s}(x))^{k-1}e^{-\bar{m}{\cal A}_{\rm s}(x)}\\\times 
\pc_{\rm s}\bigg(2^{\frac{\rho(k+t)}{W_{\rm b}}}-1,2^{\frac{\rho k}{W_{\rm a}}}-1\big| x\bigg)f_{X}(x){\rm d}x \frac{1}{\sigma_{\rm s}\sqrt{2\pi}}e^{-\frac{(t-\upsilon_{\rm s})^2}{2\sigma_{\rm s}^2}}{\rm d}t,
\end{multline}
and (iii) for average load-based partition,
\begin{multline}\label{eq::rate::cov::avg-load::sbs::PoissonN}
\pr_{\rm s}=\sum\limits_{k=1}^{\infty}\frac{\bar{m}^{k-1}}{(k-1)!}
\int\limits_{-\infty}^{\infty}\int\limits_{0}^{R-R_{\rm s}}
({\cal A}_{\rm s}(x))^{k-1}e^{-\bar{m}{\cal A}_{\rm s}(x)}\\\times 
\pc_{\rm s}\bigg(2^{\frac{\rho k\left(1+\bar{m}{\cal A}_{\rm s}(x)+\bar{m}t\right)}{W_{\rm b}(1+\bar{m}{\cal A}_{\rm s}(x))}}-1,2^{\frac{\rho k}{W_{\rm a}}}-1\big| x\bigg)\\\times f_{X}(x){\rm d}x \frac{1}{\sqrt{2\pi(n-1){\rm Var}[{\cal A}_{\rm s}(X)]}}e^{-\frac{\big(t-(n-1)\nbbE[{\cal A}_{\rm s}(X)]\big)^2}{2(n-1){\rm Var}[{\cal A}_{\rm s}(X)]}}{\rm d}t.
\end{multline}
%and $\pr_{\rm a}$ is given by \eqref{eq::rate::cov::macro::fixedN}.
\end{cor}
\begin{IEEEproof}The result can be similarly obtained from Theorem~\ref{thm::rate::cov::equal::partition} by using the PMFs of $N_{\bf x}^{\rm ABS}$, $N_{\bf x}^{\rm SBS}$, $N_{o}^{\rm ABS}$ and $N_{o}^{\rm ABS}$ from Lemmas~\ref{lemm::load::characterization::abs} and \ref{lemm::load::characterization::others}, and substituting $\nbbE[N_{\bf x}^{\rm SBS}]$ {from} Corollary~\ref{cor::mean::representative::loads} for \case~$2$. 
\end{IEEEproof}
\begin{figure*}%
\centering
\subfigure[Equal partition.]{
\label{fig::comparison::rate::cov::bw::eq}
 \includegraphics[width=.3\linewidth]{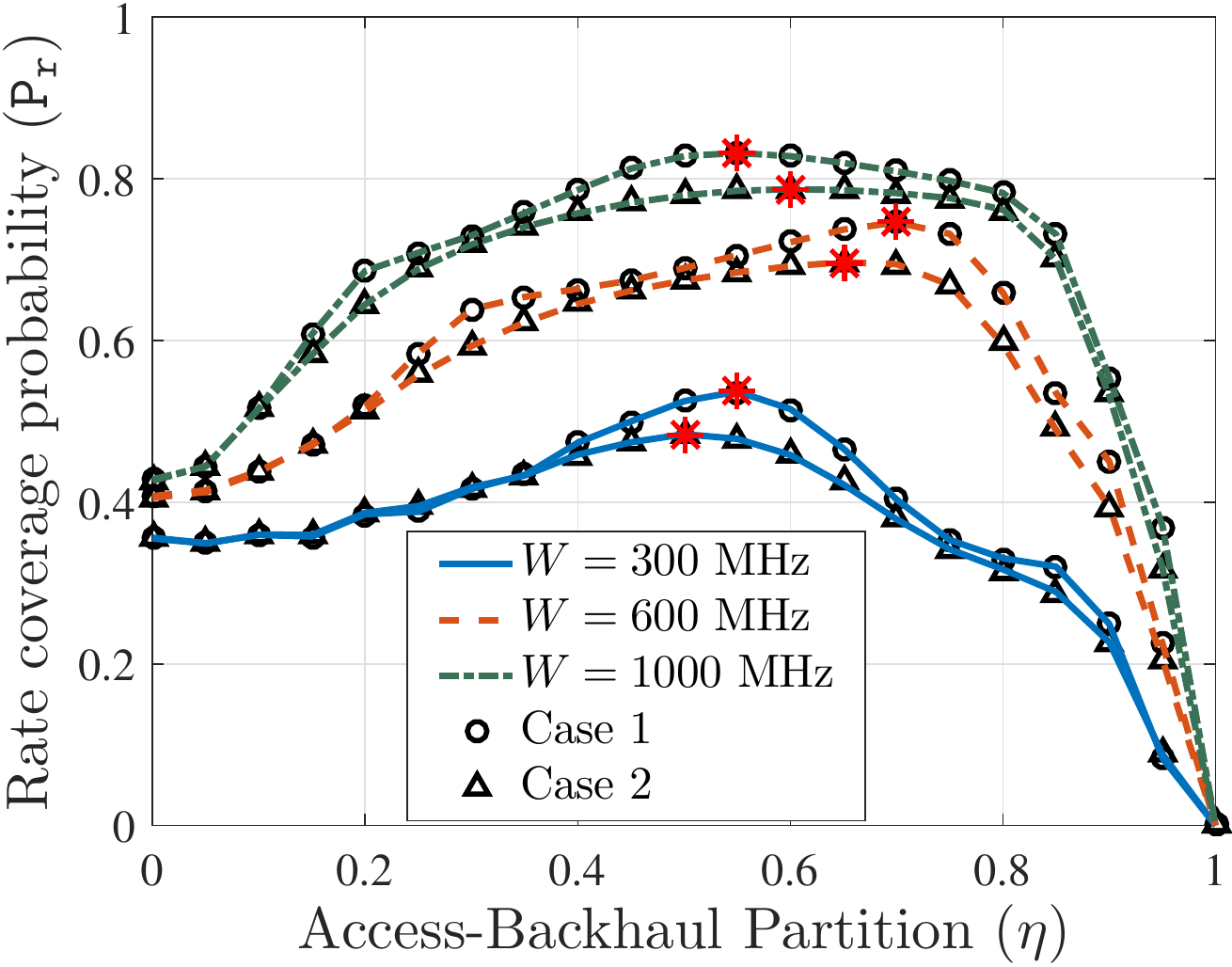}}
\hspace{8pt}%
\subfigure[Instantaneous load-based partition.]{%
\label{fig::comparison::rate::cov::bw::load}
 \includegraphics[width=.3\linewidth]{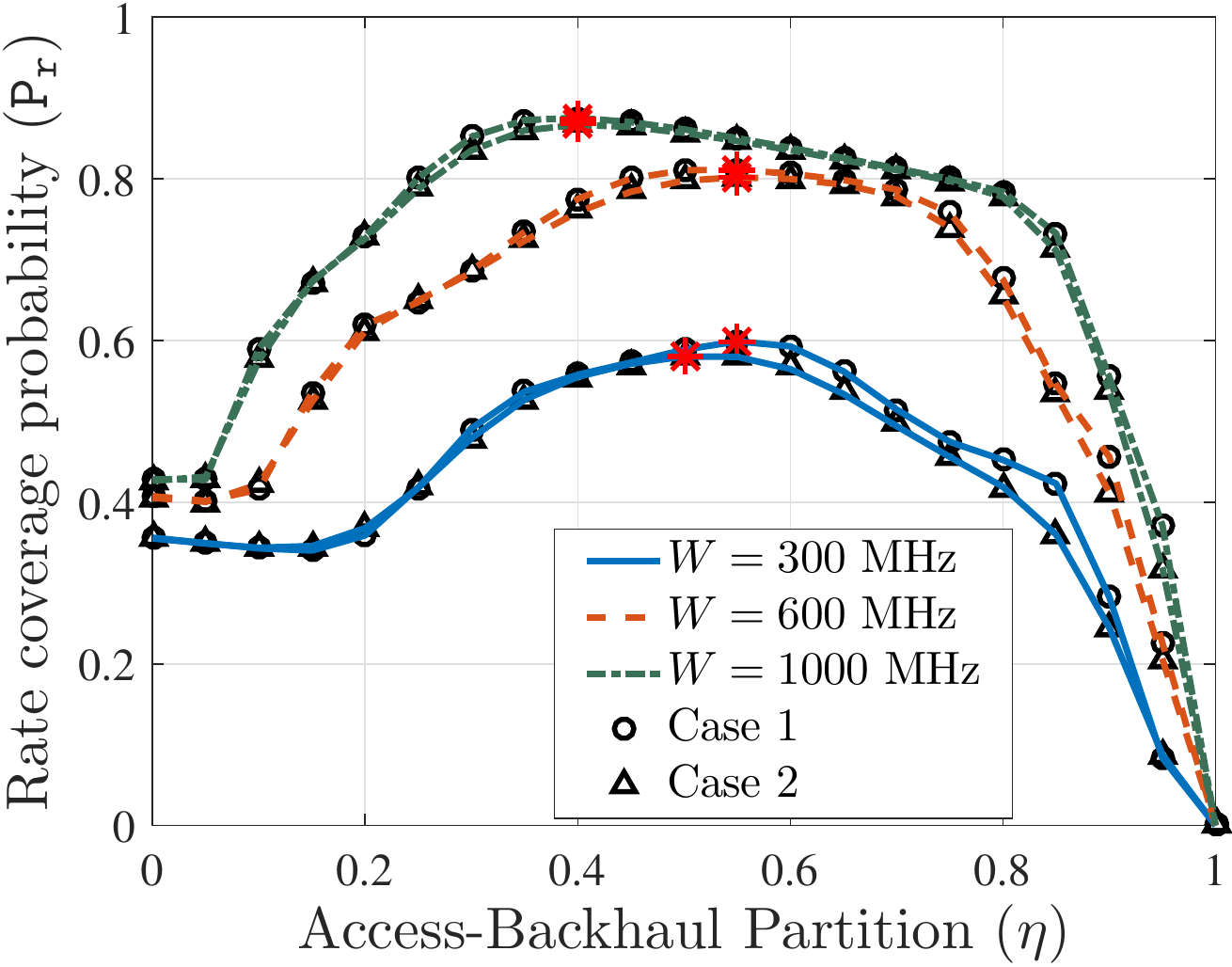}}
\subfigure[Average load-based partition.\newline]{%
\label{fig::comparison::rate::cov::bw::avg::load}
\includegraphics[width=0.3\linewidth]{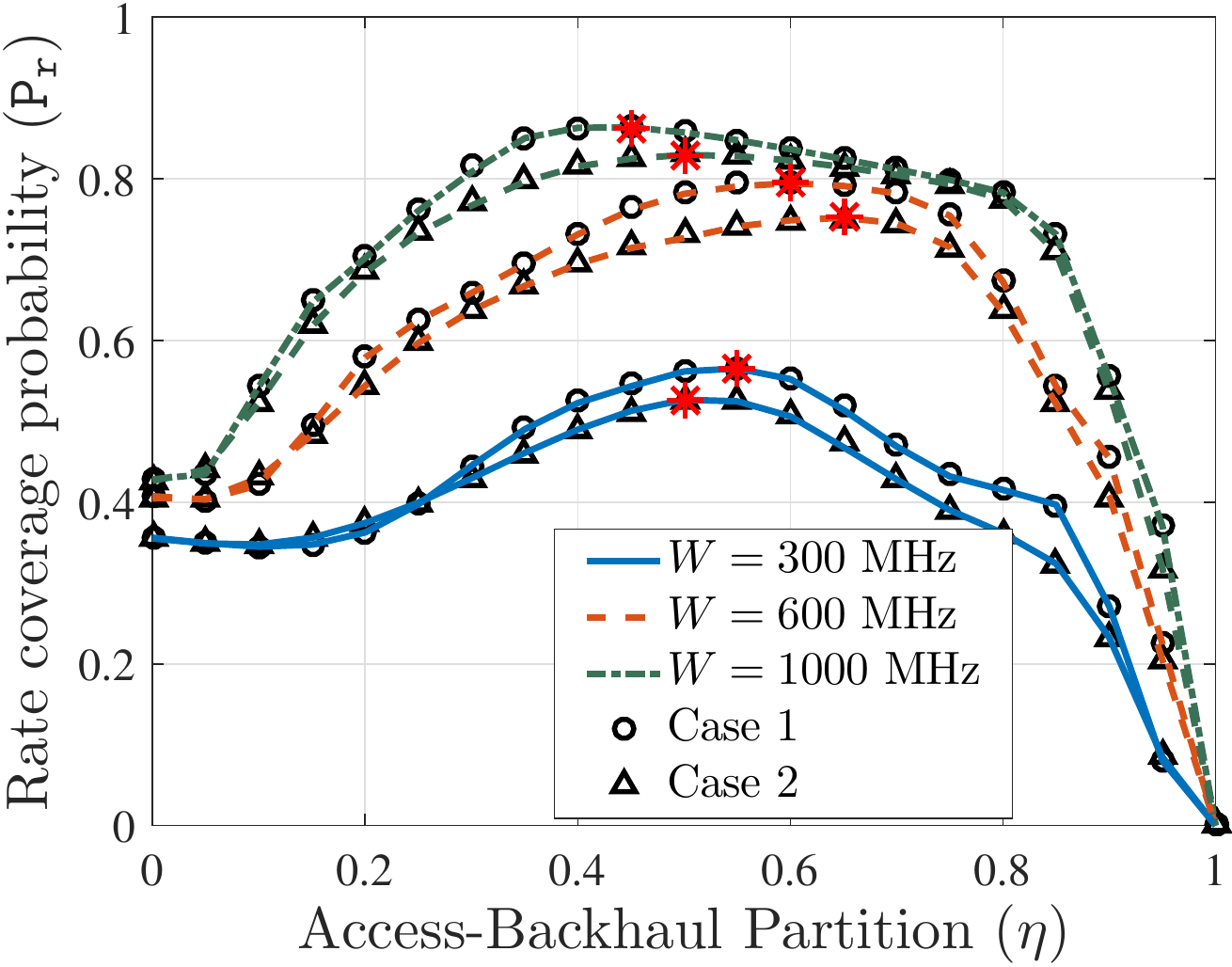}}
%\subfigure[Comparison of backhaul partition strategies for \case~$1$.]{
%\label{fig::comparison::rate::cov::bw}
%     \includegraphics[width=.22\linewidth]{./Fig/ratecov_bw_comp.pdf}%\vspace{-.7cm}
%}
\caption[Rate coverage probability  for different bandwidths ($\rho = 50$ Mbps, $n=10$) for \case s~$1$ and $2$ obtained by Corollaries~\ref{cor::rate::cov::fixedN} and \ref{cor::rate::cov::PoissonN}.]{Rate coverage probability  for different bandwidths for \case~$1$ and $2$ obtained by Corollaries~\ref{cor::rate::cov::fixedN} and \ref{cor::rate::cov::PoissonN} ($\rho = 50$ Mbps, $n=10$). {Lines and markers indicate theoretical and simulation results, respectively. Theoretical results for \case s $1$ and $2$ are obtained from Corollaries~\ref{cor::rate::cov::fixedN} and \ref{cor::rate::cov::PoissonN}, respectively.}}
\label{fig::comparison::rate::cov}%
\end{figure*}
%We will investigate the tightness of this approximation in the next Section. 
\begin{figure}
\centering
\includegraphics[scale=0.47]{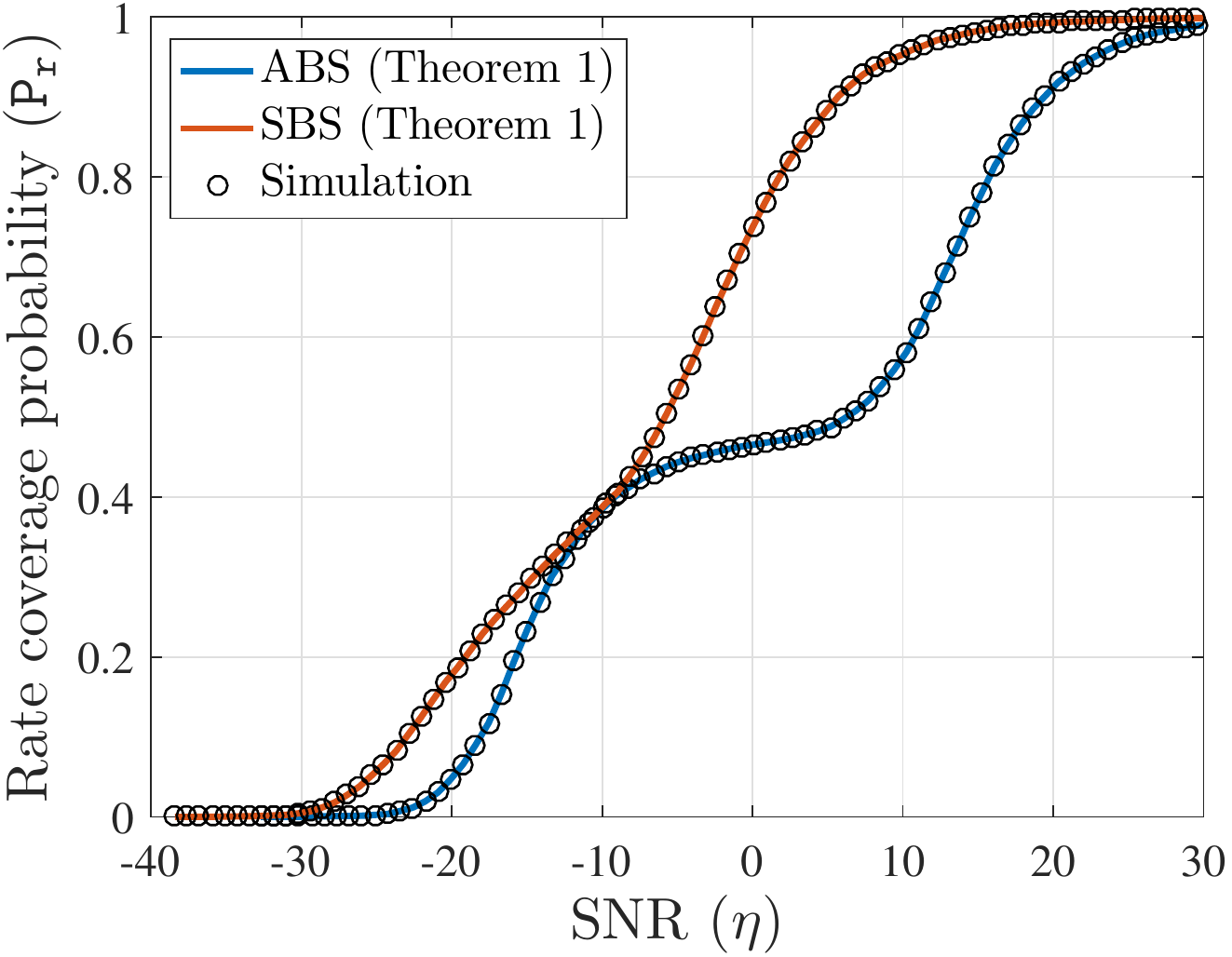}
\caption{The CDF plot of SNR from the ABS and SBS ($P_{\rm m}= 50$ dBm, $P_{\rm s} = 20$ dBm). The markers indicate empirical CDF obtained from Monte Carlo simulations.}\label{fig::snrplot}
\end{figure}
\section{Results and Discussion}\label{sec::results} 
\subsection{Trends of rate coverage}\label{subsec::trends}
In this Section we {verify the accuracy of} our analysis of rate coverage  with Monte Carlo simulations of the network model delineated in Section~\ref{sec::system::model}  with parameters listed in Table~\ref{tab::parameters}. 
 {For each simulation, the number of iterations was set to $10^6$. Since $\pr$ fundamentally {depends upon} $\snr$, we first plot the cumulative density function (CDF) of SNRs without beamforming in Fig.~\ref{fig::snrplot}, averaged over user locations. Precisely we plot $
\nbbE_{\bf x}\left[\nbbP\left(\frac{h_{\rm m} P_{\rm m}\|{\bf x}\|^{-\alpha}}{{\tt N}_0W}<\theta\right)\right] = \int_0^{R-R_{\rm s}}\pc_{\rm m}(\theta|x)f_X(x){\rm d}x
$ and  $
\nbbE_{\bf u}\left[\nbbP\left(\frac{h_{\rm s} P_{\rm s}\|{\bf u}\|^{-\alpha}}{{\tt N}_0W}<\theta\right)\right] =\pc_{\rm s}(-\infty,\theta|x)
$, where $\pc_{\rm m}$ and $\pr_{\rm s}$ were defined in Theorem~\ref{thm::coverage::probability} from  simulation and using our analytical results and observe a perfect match.} We now plot the rate coverages for different user distributions (\case s $1$ and $2$) for three different backhaul BW partition strategies  in  Figs.~\ref{fig::comparison::rate::cov::bw::eq}-\ref{fig::comparison::rate::cov::bw::avg::load}. 
 %{The markers indicate the values of $\pr$ obtained from simulation and the solid lines are drawn according to the values of $\pr$ obtained by Corollaries~\ref{cor::rate::cov::fixedN} and \ref{cor::rate::cov::PoissonN}.}  
% for equal and load-based partitions, respectively. 
Recall that one part of ABS and SBS load was approximated using CLT in Lemma~\ref{lemm::load::characterization::others} for efficient computation.  Yet, we obtain a perfect match between simulation and analysis even for $n=10$ for \case~$1$ and \case~$2$.  
  {Further, we  observe that, (i) $\pr = 0$ for $\eta=1$ since this corresponds to the extreme when no BW is given to access links, and (ii) the rate coverage is maximized  for a particular access-backhaul BW split ($\eta^*=\arg\max_{\{\eta\}}\pr$).} 
 Also note that the rate coverage trends for \case s~$1$ and $2$ are the same, although $\pr$ for \case~$1$ is slightly higher than $\pr$ of \case~$2$ since the representative cluster, on average, has more number of users in \case~$2$ than in \case~$1$ (see Corollary~\ref{cor::mean::representative::loads}). However, for space constraint, we only present the results of  \case~$1$ for subsequent discussions. 
 %\chb{In Fig.~\ref{fig::comparison::rate::from3gpp}, we compare these different backhaul BW partition strategies for a different set of transmission powers, $P_{\rm m} = 43$ dBm and $P_{\rm s} = 24$ dBm which are obtained from 3GPP specification~\cite{access2010further} and observe the similar trend as in Fig.~\ref{fig::comparison::rate::cov::bw}. Note that $\eta^*$ is also sensitive to the choice of transmission powers.} 

%\begin{figure*}
\begin{figure}
\centering
    \includegraphics[width=.7\linewidth]{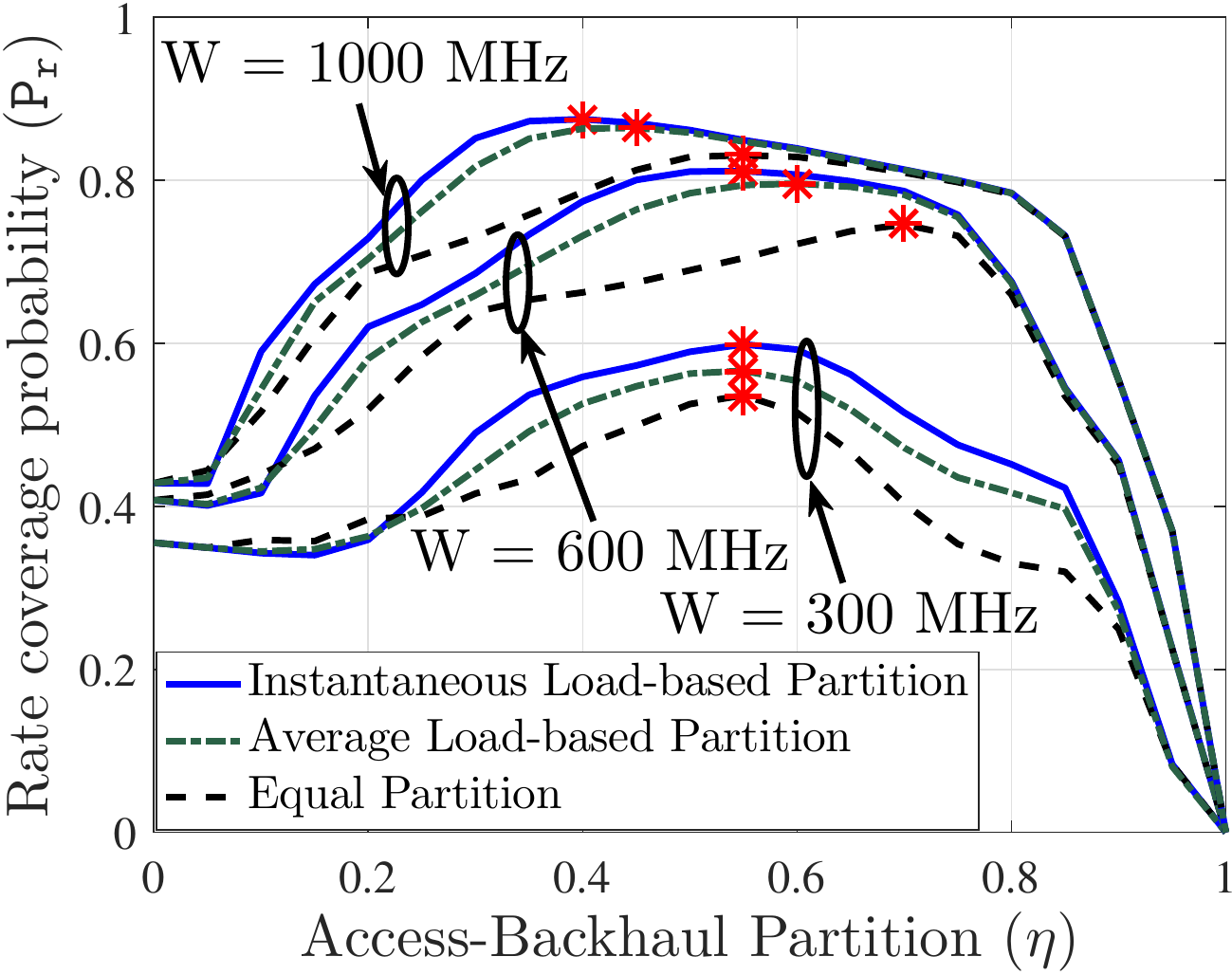}%\vspace{-.7cm}
%}
\caption{Comparison of backhaul partition strategies for \case~$1$ ($\rho = 50$ Mbps, $n=10$).}
\label{fig::comparison::rate::cov::bw}
\end{figure}
\begin{figure}%{0.49\linewidth}
%       \begin{figure}[H]
\centering
               \includegraphics[width=.7\linewidth]{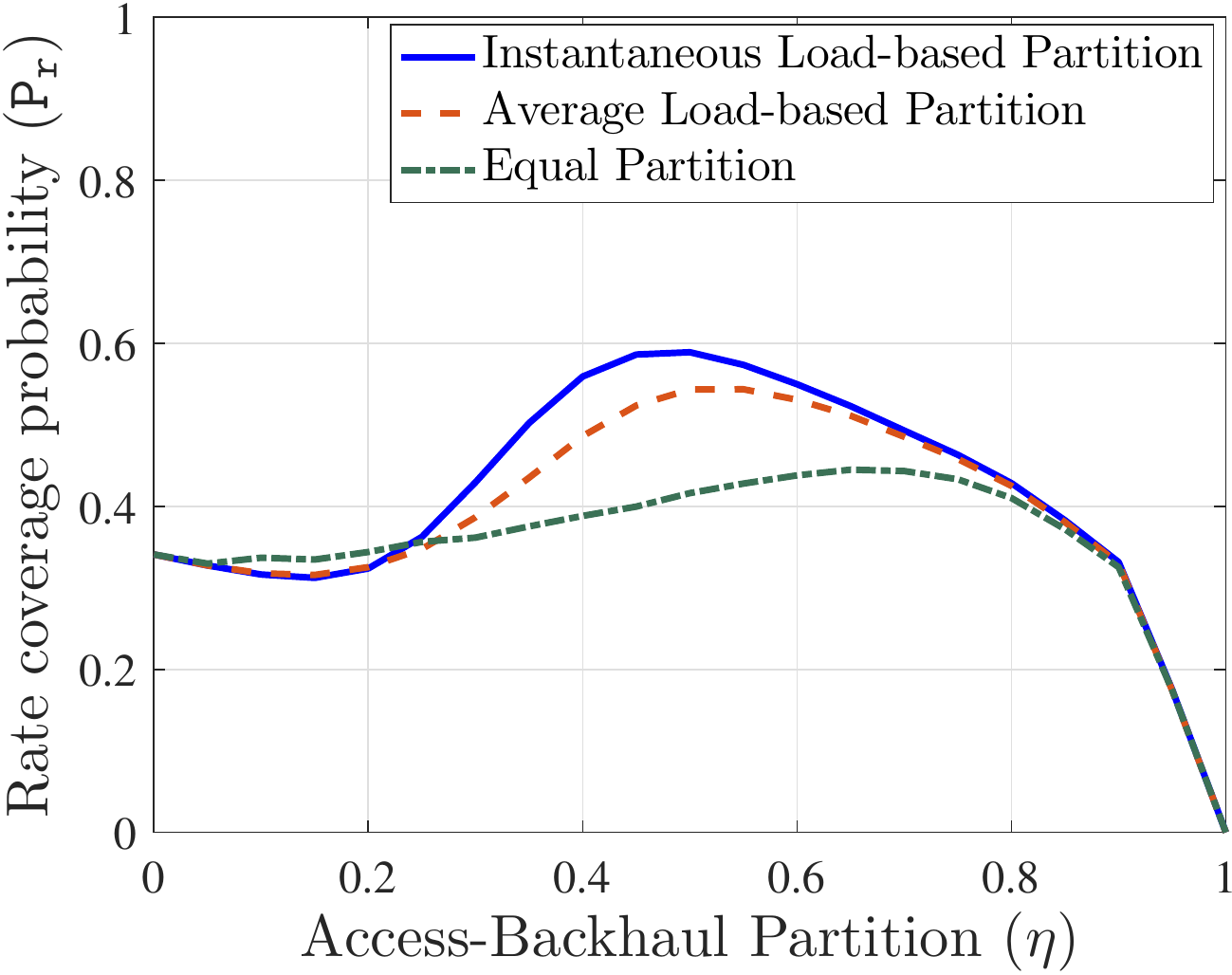}%\vspace{-.7cm}
\caption{Comparision of backhaul BW parition strategies  ($\rho = 50$ Mbps, $n=10$, $W = 600$ MHz) for \case~$1$ and $\mu= 30$ m. {The results are obtained from Corollary~\ref{cor::rate::cov::fixedN}.}\newline\newline}\label{fig::comparison::rate::cov::bw::lowmu}
       \end{figure}       
       \begin{figure}
       \centering
             \includegraphics[width=.7\linewidth]{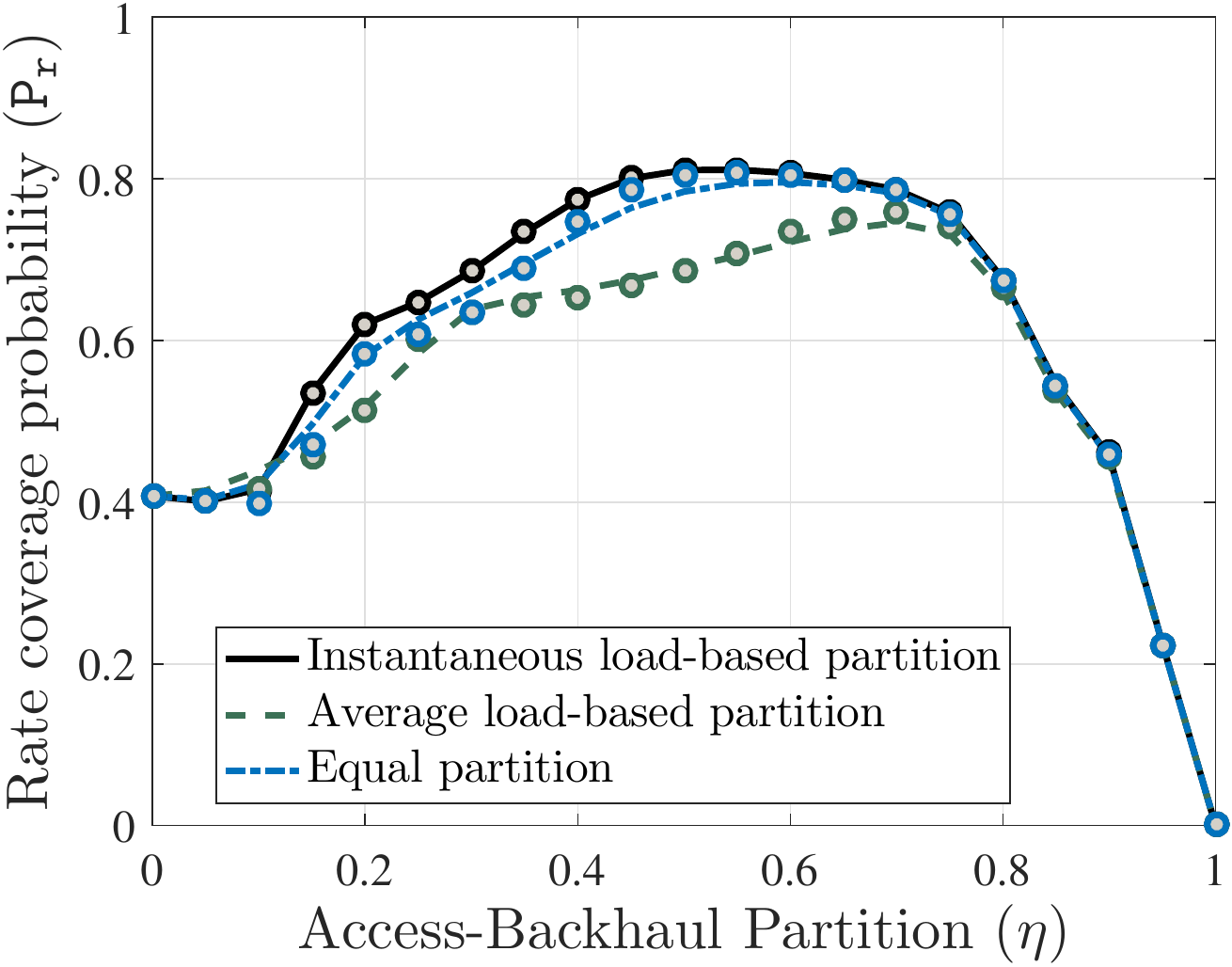}%\vspace{-.7cm}
\caption{Comparison of the exact expression (Corollary~\ref{cor::rate::cov::fixedN}) and approximate expression (Lemma~\ref{lemm::approximation}) of rate coverage probability for \case~$1$ ($\rho = 50$ Mbps, $W = 600$ MHz, $n=10$).  Lines and markers indicate exact and approximate results, respectively. }\label{fig::comparison::rate::approx::fixedN}

\end{figure}
%\end{figure*}

%\begin{figure}
%\centering
%\begin{minipage}{0.49\linewidth}
%%       \begin{figure}[H]
%               \includegraphics[width=.8\linewidth]{./Fig/low_mu.pdf}%\vspace{-.7cm}
%\caption{Comparision of backhaul BW parition strategies  ($\rho = 50$ Mbps, $n=10$, $W = 600$ MHz) for \case~$1$ and $\mu= 30$ m. {The results are obtained from Corollary~\ref{cor::rate::cov::fixedN}.}\newline}\label{fig::comparison::rate::cov::bw::lowmu}
%%       \end{figure}
%\end{minipage}
%  \begin{minipage}{0.49\linewidth}
% %         \begin{figure}[H]
%             \includegraphics[width=.8\linewidth]{./Fig/ratecov_approx_fixed_user.pdf}%\vspace{-.7cm}
%\caption{Comparison of the exact expression (Corollary~\ref{cor::rate::cov::fixedN}) and approximate expression (Lemma~\ref{lemm::approximation}) of rate coverage probability for \case~$1$ ($\rho = 50$ Mbps, $W = 600$ MHz, $n=10$).  Lines and markers indicate exact and approximate results, respectively. }\label{fig::comparison::rate::approx::fixedN}
%% \end{figure}
% \end{minipage}
%\end{figure}   
\subsubsection{Comparison of backhaul BW partition strategies}
In Fig.~\ref{fig::comparison::rate::cov::bw}, we overlay $\pr$ for three different backhaul BW partition strategies.   We observe that the maximum rate coverage, $\pr^* = \pr(\eta^*)$ (marked as `*' in the figures) for instantaneous load-based partition dominates $\pr^*$  in average load-based partition, {and}  $\pr^*$  in average load-based partition dominates $\pr^*$ in equal partition. {Also note that $\eta^*$ is different for different combination of BW partition strategy and $W$. We  further compared these three strategies in a high blocking environment  in Fig.~\ref{fig::comparison::rate::cov::bw::lowmu} by setting $\mu = 30$ m and observe the same ordering of performance of the three strategies. As expected, $\pr$ is in general lower for this case.}  That said, it should be kept in mind that  instantaneous load-based partition requires more frequent feedback of the load information from the SBSs and hence  has the  highest signaling overhead among the three strategies.  The average load-based partition requires comparatively less signaling overhead since  it does not require frequent feedback. On the other hand, equal partition does not have this overhead at all.  This motivates an interesting  performance-complexity trade-off for the design of cellular networks with IAB.

 \begin{figure}
\centering
         \includegraphics[width=.7\linewidth]{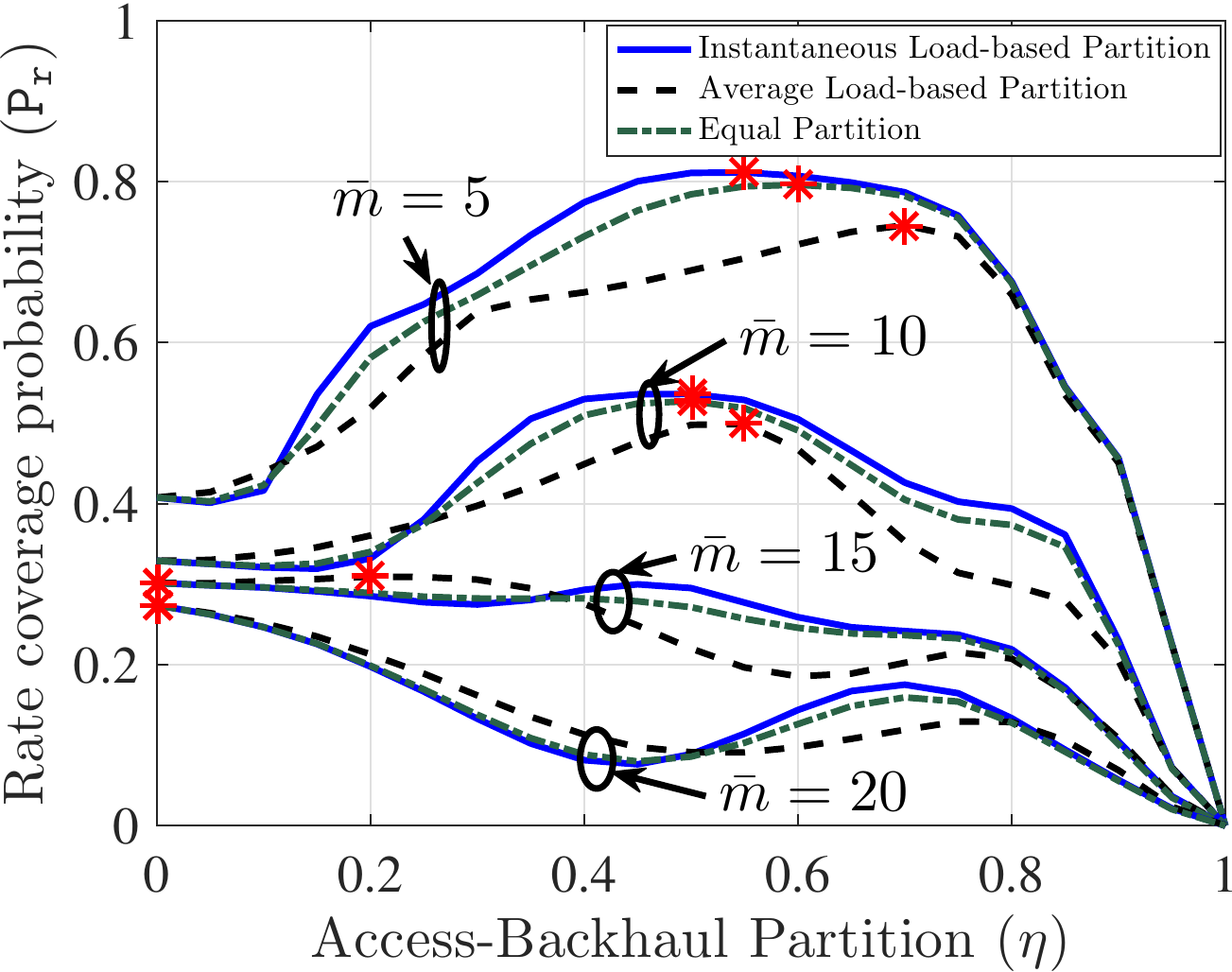}%%\vspace{-.7cm}
\caption{Rate coverage for different numbers of users per hotspot for \case~$1$ ($W = 600$ MHz, $\rho = 50$ Mbps). The values of $\pr$ are computed using Lemma~\ref{lemm::approximation}.}\label{fig::comparison::rate::cov::users}
       \end{figure}
 \begin{figure}
 \centering
 \includegraphics[width=.7\linewidth ]{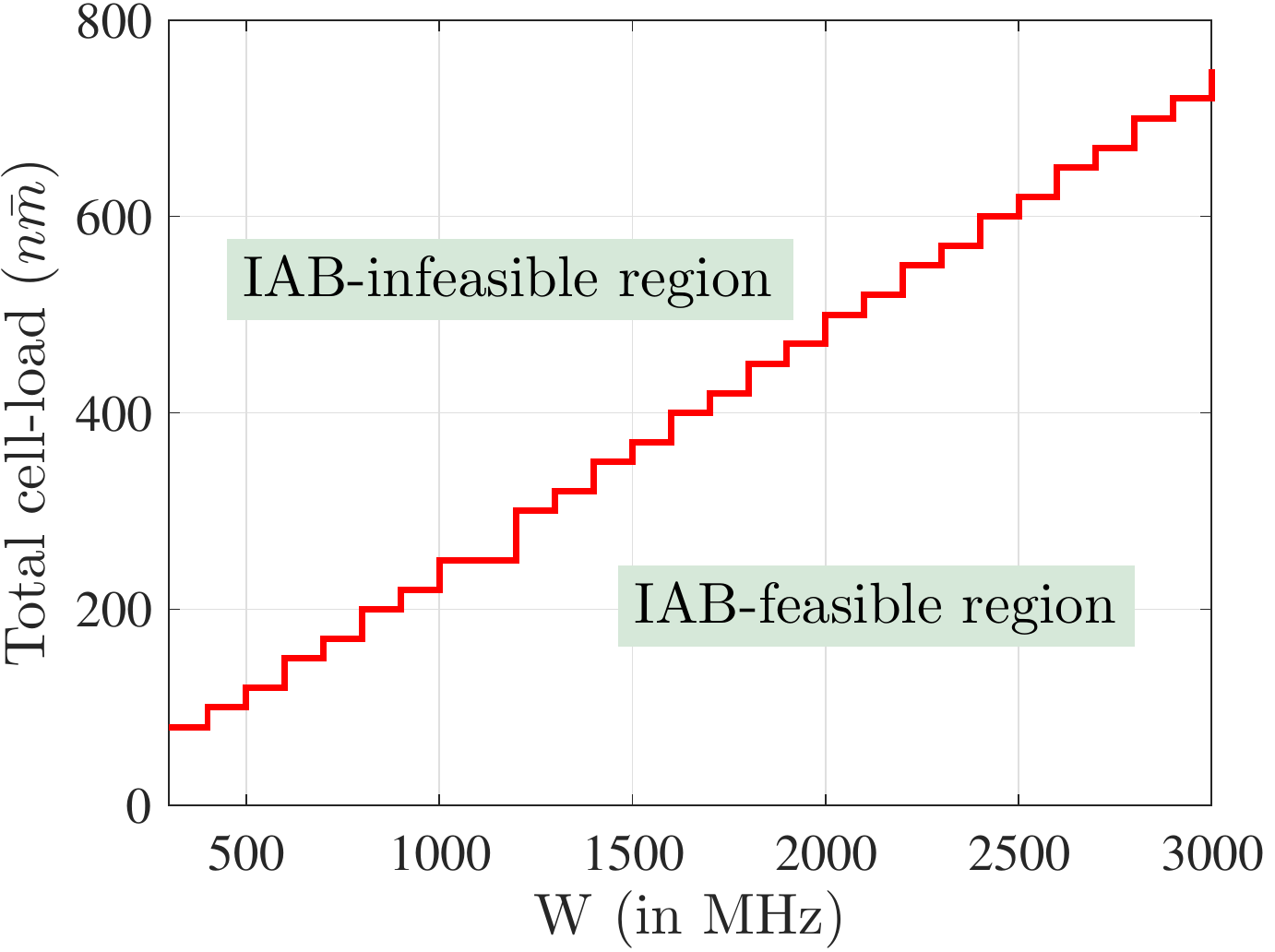}
 %\vspace{-.7cm}
   \caption{Total  cell-load upto which IAB-enabled network outperforms macro-only network (for instantaneous load-based partition).\newline}\label{fig::total::critical::load}
 \end{figure}
\subsubsection{Effect of system BW}
We observe the effect of increasing system BW on rate coverage in Fig~\ref{fig::comparison::rate::cov::bw}. As expected, $\pr$ increases as $W$ increases. However, the increment of $\pr^*$ saturates for very high values of $W$ since high noise power {degrades} the link spectral efficiency. Another interesting observation is that starting from $\eta = 0$ to $\eta^*$, $\pr$ does not increase monotonically. This is due to the fact that sufficient BW needs to be steered from access to backhaul so that the network with IAB performs better than the macro-only network (corresponding to $\eta = 0$).
\subsubsection{Accuracy of approximation}We now plot $\pr$  obtained by  the approximations in Lemma~\ref{lemm::approximation} in Fig.~\ref{fig::comparison::rate::approx::fixedN}. It is observed that the approximation is surprisingly close to the exact values of $\pr$ obtained by Corollary~\ref{cor::rate::cov::fixedN}. 
Motivated by the tightness of the approximation, we proceed with the easy-to-compute expressions of $\pr$ obtained by Lemma~\ref{lemm::approximation} instead of the exact expressions (Corollary~\ref{cor::rate::cov::fixedN}) for the metrics evaluated in the sequel, namely, critical load, median rate, and $5^{th}$ percentile rate. {It is important to note that each numerical evaluation of  these metrics requires high number of computations of  $\pr$ and is  highly inefficient if $\pr$ is computed  by simulation,  which further highlights the importance of analytical expressions derived in this paper.} 
\subsection{Critical load}
We plot the variation of $\pr$ with $\bar{m}$ in Fig.~\ref{fig::comparison::rate::cov::users}. We observe that as $\bar{m}$ increases,   more number of users share the BW and as a result, $\pr$ decreases. However, the optimality of $\pr$ completely disappears for very large value of $\bar{m}$ ($10<\bar{m}<20$ in this case). This implies that for given BW $W$ there exists a {\em critical total cell-load} ($n\bar{m}$) beyond which the gain obtained by the IAB architecture  completely disappears. 
{Observing Fig.~\ref{fig::total::critical::load},  we find that  the critical total cell-load  varies linearly with the system BW.} 
 The reason of the existence of the critical total cell-load can be intuitively explained as follows. Recall that the SBS rate ${\cal R}_{\rm a}^{SBS}$ was limited by the backhaul constraint ${\cal R}_{\rm b}^{\rm ABS}/N_{\bf x}^{\rm SBS}$. 
When $\bar{m}$ is high, $N_{\bf x}^{\rm SBS}$ is also high and this puts stringent backhaul constraint on  ${\cal R}_{\rm a}^{\rm SBS}$. Hence, an ABS can serve more users by direct macro-links at the target rate instead of allocating any backhaul partition.% directed by any of the three strategies. 
%Consider two hotspots ($n=2$) at ${\bf x}_1$ and ${\bf x}_2$, such that ${\bf x}_1$ is near the cell center  and ${\bf x}_2$ is  near the cell edge (i.e., $0<x_1<x_2<R-R_{\rm s}$). Under this situation, we can observe the following:  (i) the users of the hotspot  at ${\bf x}_1$ will experience better {ABS rate} ($\pr_{\rm m}$ compared to the users of the hotspot  at ${\bf x}_2$ because of the smaller ABS link distance, (ii) more number of users connects to the SBS at the hotspot at ${\bf x}_2$, i.e., $N_{{\bf x}_2}^{\rm SBS}>N_{{\bf x}_1}^{\rm SBS}$. When BW is steered from access to backhaul by increasing $\eta$, SBS links can achieve higher throughput. As a result,   $\pr$ increases when the increase in $\pr_{\rm s}$ is dominated by the decrease in $\pr_{\rm m}$.   But, note that the SBS rate at the hotspot  at ${\bf x}_i$ is upperbounded by ${\cal R}_{\rm b}/N^{\rm SBS}_{{\bf x}_i}$ and ${\cal R}_{\rm b}$ is also limited by the $\snr$ observed at ${\bf x}_i$. When $\bar{m}$ becomes very high, the improvement of $\pr_{\rm s}$ for the hotspot at ${\bf x}_2$ is not prominent at the expense of the decrease of $\pr_{\rm m}$. Hence, $\pr $ is maximized at $\eta=0$.} 
%\vspace{-1cm}
\begin{figure}
\centering
\includegraphics[width=.7\linewidth]{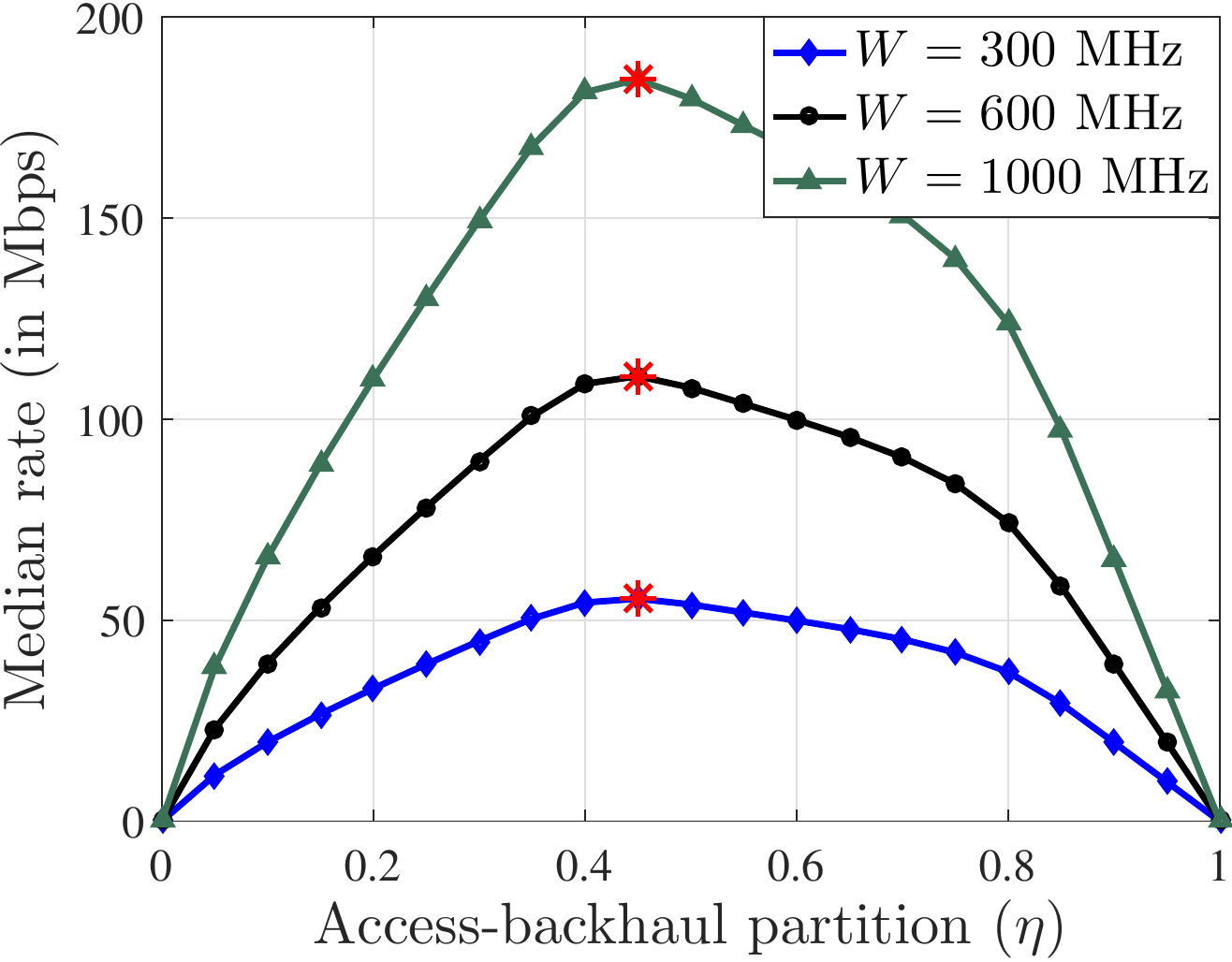}
\caption{Median rate for \case~$1$ for instantaneous load-based partition ($n=10$).}\label{fig::median::rate}
\end{figure}
\begin{figure}
\centering
\includegraphics[width=.7\linewidth]{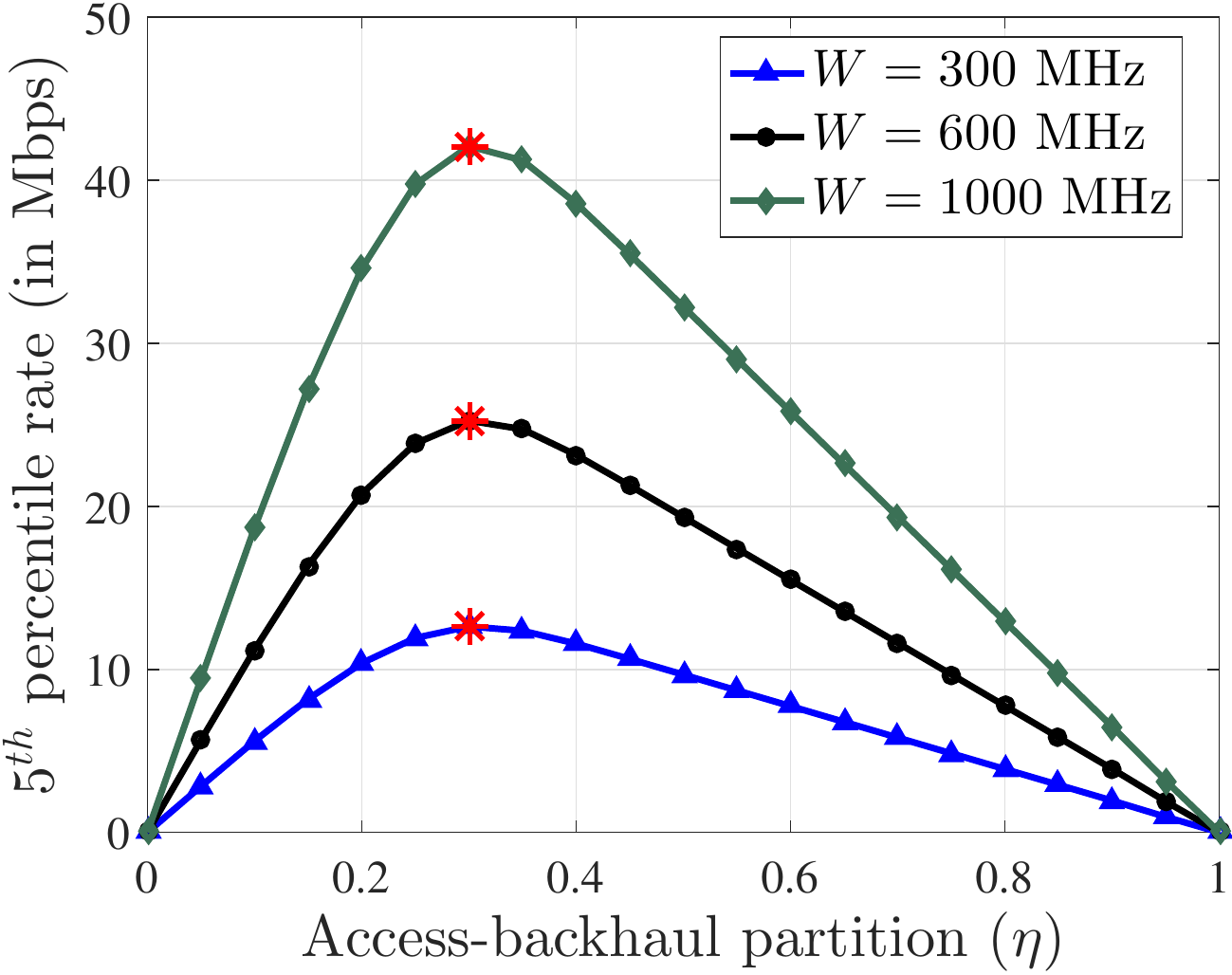}
\caption{$5^{th}$ percentile rate for \case~$1$ for instantaneous load-based partition ($n=10$).}\label{fig::5thpercentile::rate}
\end{figure}
\begin{figure}
\centering
\includegraphics[width=.7\linewidth]{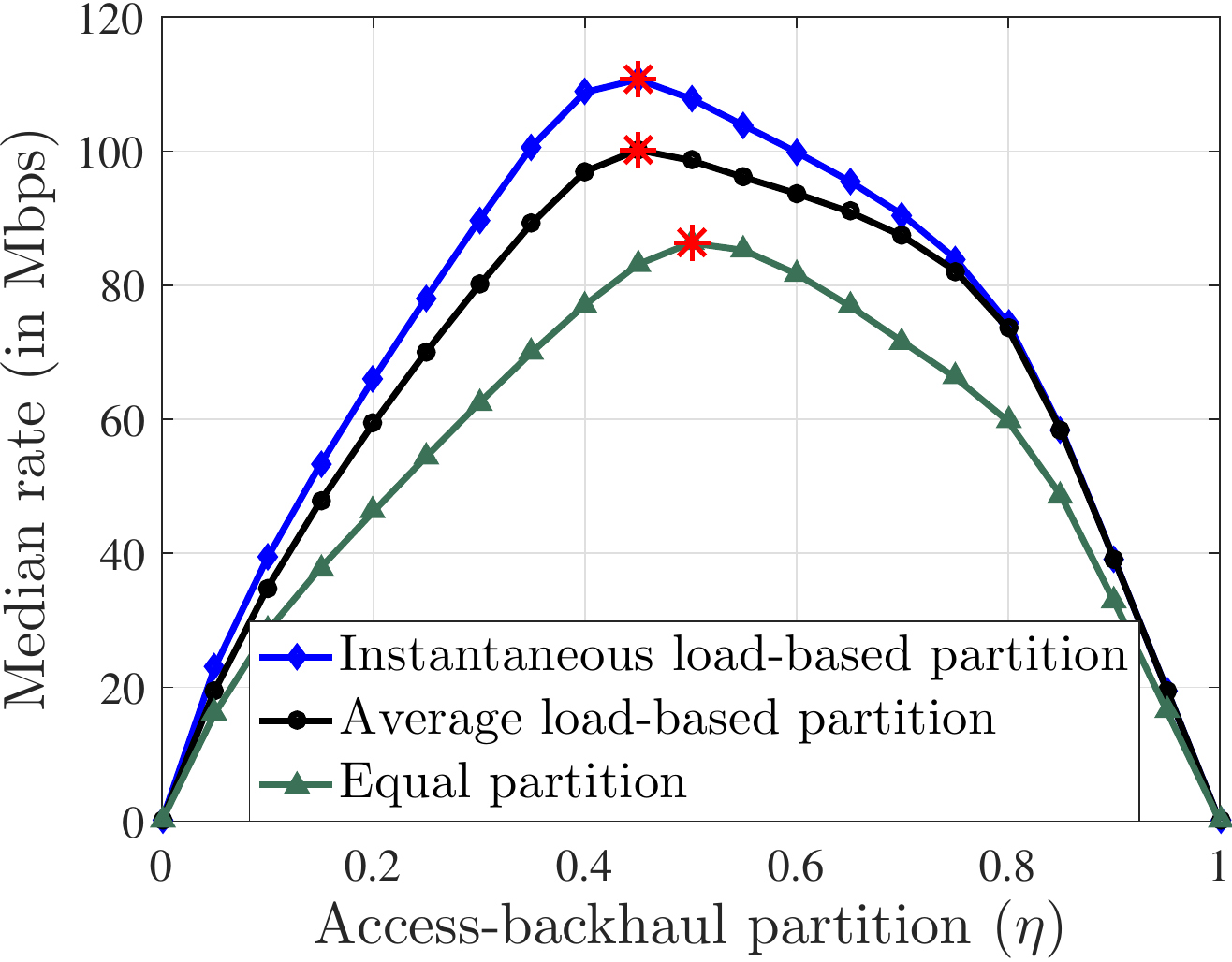}
\caption{{Median rate for \case~$1$ for different backhaul BW partition strategies ($n=10$, $W = 600$ MHz).}\newline}\label{fig::median::rate::comp}
\end{figure}
\begin{figure}
\centering
\includegraphics[width=.7\linewidth]{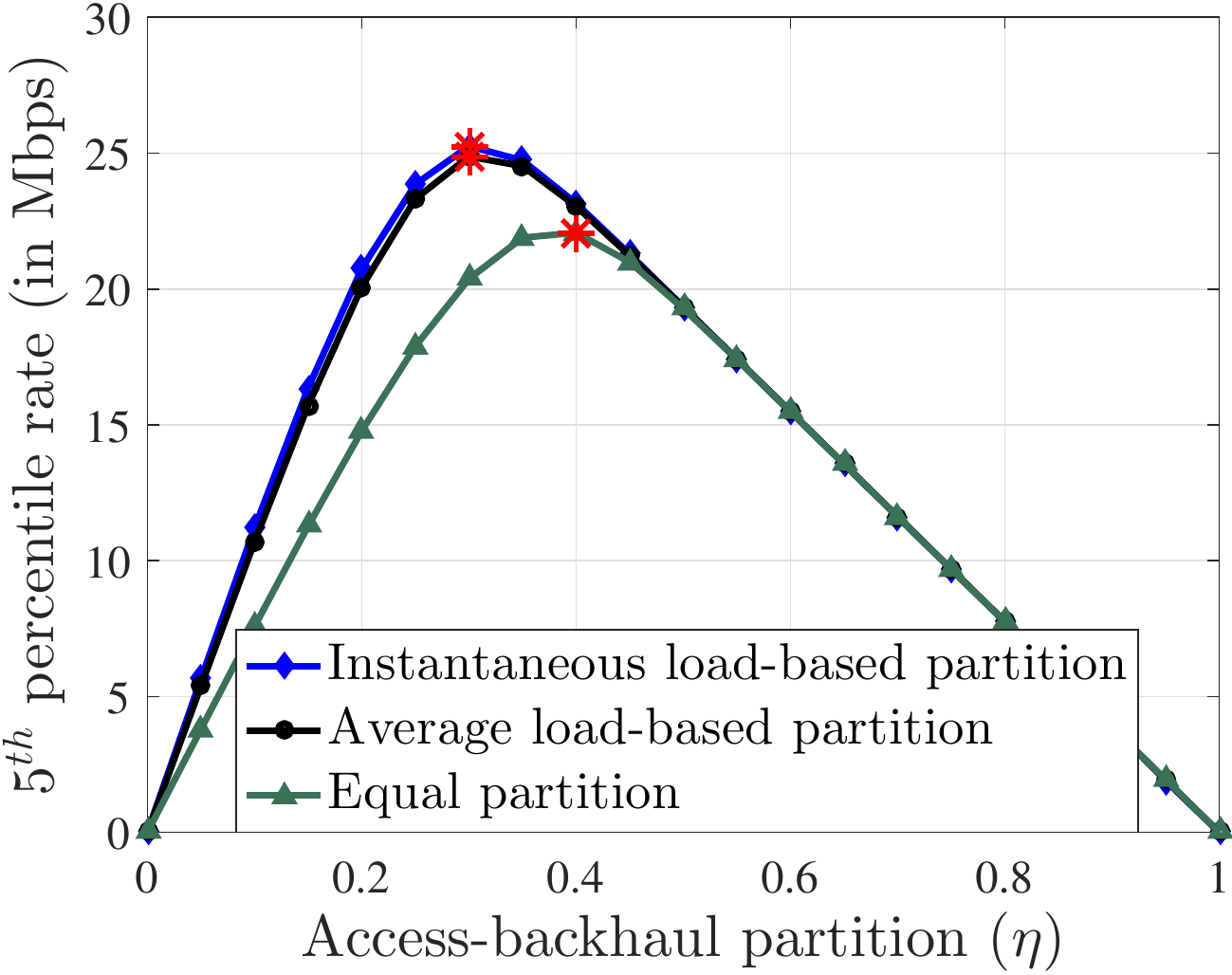}
\caption{{$5^{th}$ percentile rate for \case~$1$  for different backhaul BW partition strategies ($n=10$, $W = 600$ MHz).}}\label{fig::5thpercentile::rate::comp}
\end{figure}
\subsection{Median and $5^{th}$ percentile rates}
We now shift our attention to two more performance metrics of interest,  the median ($50^{th}$  percentile)   and $5^{th}$ percentile rate, which are denoted as $\rho_{50}$ and $\rho_{95}$, respectively. These rates are defined as the values where the rate CDF attains $0.5$ and $0.05$, respectively, i.e.,  $\pr = 0.5$ at $\rho=\rho_{50}$ and   $\pr = 0.95$ at $\rho=\rho_{95}$.      
Figs.~\ref{fig::median::rate} and \ref{fig::5thpercentile::rate} illustrate  $\rho_{50}$ and $\rho_{95}$ respectively for different $W$. {We first observe that for a given $\eta$, these rates increase linearly with $W$. This is because of the fact that in all the expressions of rate coverage, $\rho$ and $W$ appear as a ratio ($\rho/W$). Thus, once we find  a desired rate coverage at a particular $\rho$ for a given $W$, same rate coverage will be observed for $kW$ at target data rate $k\rho$ (where $k$ is a positive constant).} Further, we notice that the median rate is relatively flat around the maximum compared to the $5^{th}$ percentile rate. Also, the optimal $\eta$ does not vary significantly (stays close to 0.4 in our setup) for median and $5^{th}$ percentile rates. In Figs.~\ref{fig::median::rate::comp} and \ref{fig::5thpercentile::rate::comp}, we have compared the three backhaul BW partition strategies in terms of these two rates. As expected, the ordering in performance is similar to the one observed for $\pr^*$. Interestingly, from Fig.~\ref{fig::5thpercentile::rate::comp}, it appears that the average and instantaneous load-based partition policies have almost similar performance in terms of  $5^{th}$ percentile rate.  {This is because of the fact that  $\rho_{95}$ is towards the tail of the rate distribution which is not significantly affected by  difference between instantaneous or average load.  However, the performance gap becomes prominent once  median rate is considered.}  
%\begin{figure}
%\centering
%\includegraphics[scale=0.5]{./Fig/ratecov_bw.pdf}
%\caption{Rate coverage probability for load-based partition for different bandwidths ($\rho = 50$ Mbps).}\label{fig::comparison::rate::cov::bw}
%\end{figure}
%\begin{figure}
%\centering
%\includegraphics[scale=0.5]{./Fig/ratecov_partition.pdf}
%\caption{Comparision of equal and load-based parition policies ($W = 300$ MHz, $\rho = 50$ Mbps).}\label{fig::comparison::rate::cov::bw}
%\end{figure}
%\begin{figure}
%\centering
%\includegraphics[scale=0.4]{./Fig/ratecov_users.pdf}
%\caption{Rate coverage for different numbers of users per hotspot ($W = 600$ MHz, $\rho = 50$ Mbps).}\label{fig::comparison::rate::cov::users}
%\end{figure}
\section{Conclusion}\label{sec::conclusion}
In this paper, we proposed the first 3GPP-inspired analytical framework for two-tier mm-wave
HetNets with IAB and investigated three backhaul BW partition strategies. In particular, our
model was inspired by the spatial configurations of the users and BSs considered in 3GPP simulation models of mm-wave IAB, where the SBSs are deployed at the centers of user hotspots. Under the assumption
that the mm-wave communication is noise limited, we evaluated the downlink rate coverage
probability. As a key intermediate step, we characterized the PMFs of load on the ABS and SBS
for two different user distributions per hotspot. Our analysis leads to two important system-level
insights: (i) for three performance metrics namely, downlink rate coverage probability, median rate,
and $5^{th}$ percentile rate, the existence of the optimal access-backhaul bandwidth partition splits
for which the metrics are maximized, and (ii) maximum total cell-load that can be supported
using the IAB architecture. 

{
This work has numerous extensions. From the modeling perspective, although the noise power dominates interference power  in most of the operating regimes in  mm-wave networks,  it is important to consider  interference from the ABS and  the SBSs in the access links which may not be negligible in {a very} dense setup.  For the analysis of rate coverage in this case, one needs to evaluate  the joint distribution of  interference 
 and load which will be spatially coupled  by the locations of the SBSs. 
 Built on this baseline model of IAB, one can also study the spatial  multiplexing gains on the resource allocation obtained by massive multi-user-multiple input-multiple-output (MU-MIMO) transmissions in the downlink. Further,  
this framework can be used to study IAB-enabled cellular networks where the control signalling for cell association is also performed in mm-wave. Under this setup, the problems of mm-wave beam sweeping and corresponding cell association delays can be studied. Another useful extension of this work is to consider an additional {set of} of users having open access to all the SBSs, while in this paper we only considered users with closed-access to the SBSs at the hotspot center.  
 From the stochastic geometry perspective, it will be useful to develop analytical generative models for correlated blocking of the mm-wave signals making these analytical models more accurate. As said earlier, this work also {lays} the foundations of analytical characterization of cell-load in the 3GPP-inspired HetNet models.  Using the fundamentals of this load modeling approach, one can further design optimal bias-factors depending on the volume of cell load at each hotspot and improve the per-user rates. Finally, the analysis can be extended to design delay sensitive routing/scheduling policies which is also a relevant research directions for IAB-enabled networks. 
}
%
%
%analytical frameworks can be developed for other relevant problems whose studies  extensively rely on  the 3GPP simulation models of HetNets, such as user mobility and hand-off rates.  
\appendix
\subsection{Proof of Theorem~\ref{thm::coverage::probability}}
\label{app::coverage::probability}
Conditioned on the location of the typical user at ${\bf u}=(u,\xi)$ and its hotspot center at ${\bf x}$, $\pc_{\rm s}(\theta_1,\theta_2|{ x}) =$
%{we have to pay more attention in retaining or relaxing the boldface notation. Have to clarify the sampling policies and equivalence (?).}
\begin{align*}
& \nbbP(\snr^{\rm SBS}_{\rm a}({ u})>\theta_2, \snr_{\rm b}({ x})>\theta_{1},{\cal E}=1|x)=\\
&\nbbP(\snr^{\rm SBS}_{\rm a}({ u})>\theta_2,{\cal E}=1|x)\nbbP(\snr_{\rm b}({ x})>\theta_{1}|x)\\
&\myeq{a}\nbbE\bigg[\bigg(p({u}) \nbbP{\bigg(\frac{P_{\rm s}G\beta^{-1} h_{\rm s(L)} u^{-\alpha_L}}{{\tt N}_0W}>\theta_2\bigg)}\\& +(1-p({u})) \nbbP{\bigg(\frac{P_{\rm s}\beta^{-1} Gh_{\rm s(NL)} u^{-\alpha_{NL}}}{{\tt N}_0W}>\theta_2\bigg)}\bigg)\times\\&{\bf 1}\bigg(u\in{\bigg(0,{\frac{xk_p\big(\sqrt{1-k_p^2\sin^2\xi}+k_p\cos\xi\big)}{1-k_p^2}}\bigg)}, \xi\in(0,2\pi]\bigg)\bigg|x\bigg]\\&\times
\bigg(p({x}) \nbbP\bigg(\frac{P_{\rm m}\beta^{-1} G^2 h_{\rm b(L)} x^{-\alpha_L}}{{\tt N}_0W}>\theta_1|x\bigg)\\& +(1-p({x}))
\nbbP\bigg(\frac{P_{\rm m}\beta^{-1} G^2h_{\rm b(NL)} x^{-\alpha_{NL}}}{{\tt N}_0W}>\theta_1|x\bigg)\bigg).
\end{align*}
 Here (a) follows from step (a) in the proof of Lemma~\ref{lemm::association::sbs}. The final form is obtained by evaluating the expectation with respect to $u$ and $\xi$. We can similarly obtain  $\pc_{\rm m}(\theta_3|{\bf x})= $
 \begin{align*}
&\nbbP(\snr^{\rm ABS}_{\rm a}({ {\bf x}+{\bf u}})>\theta_3, {\cal E}=0|x)  \\
&= \nbbE\bigg[p(\sqrt{x^2+u^2+2xu\cos\xi})\\
&\times \nbbP{\bigg(\frac{P_{\rm m}G\beta^{-1} h_{\rm m(L)} (x^2+u^2+2xu\cos\xi)^{-\frac{\alpha_L}{2}}}{{\tt N}_0W}>\theta_2\bigg)} \\
&+(1-p(\sqrt{x^2+u^2+2xu\cos\xi}))\times \nbbP\bigg(\frac{P_{\rm m}\beta^{-1} Gh_{\rm m(NL)}}{{\tt N}_0W} \\
&\times ({x^2+u^2+2xu\cos\xi})^{-\frac{\alpha_{NL}}{2}}>\theta_2\bigg)\bigg|x\bigg],
 \end{align*}
followed by deconditioning over $u$ and $\xi$.
\subsection{Proof of Lemma~\ref{lemm::load::characterization::abs}}
\label{app::load::characterization::abs}
Conditioned on the fact that the representative hotspot is centered at $\bf x$ and the typical user connects to the ABS, $N_{\bf x}^{\rm ABS} =1+ N_{\bf x}^{\rm ABS,o}$, where $N_{\bf x}^{\rm ABS,o}$  is the load due to  rest of the users in the  representative hotspot connecting to the ABS, where, $N_{\bf x}^{\rm ABS,o} =$
\begin{align*}
\begin{cases}\nbbE[\sum_{j=1}^{{N}_{\bf x}-1} \mathbf{1}({P_{\rm m}\|{\bf x}+{\bf u}_j\|^{-\alpha}>P_{\rm s}\|{\bf u}_j\|^{-\alpha}})|{\bf x}]&\\&\text{for \case~$1$,}\\
\nbbE[\sum_{j=1}^{{N}_{\bf x}} \mathbf{1}({P_{\rm m}\|{\bf x}+{\bf u}_j\|^{-\alpha}>P_{\rm s}\|{\bf u}_j\|^{-\alpha}})|{\bf x}]&\\&\text{for \case~$2$.}
\end{cases}
\end{align*}
{Note that the difference between the above two expressions is the upper bound of the summation. Recall that, $N_{\bf x}=N_{{\bf x}_{n}}=\bar{m}$ for \case~$1$, and $N_{\bf x}=N_{{\bf x}_n}+1$  for \case~$2$ (by Remark~\ref{rem::typical}). Hence, the number of other users except the typical user in the representative cluster is $N_{\bf x}-1$ for \case~$1$ and $N_{\bf x}$ for \case~$2  $. }
 For \case~$1$,  the conditional moment generating function (MGF) of $ N_{\bf x}^{\rm ABS,o}$ is: 
\begin{align*} 
&\nbbE[e^{sN_{\bf x}^{\rm ABS,o}}|{\bf x}] =\nbbE\bigg[\prod\limits_{j=1}^{\bar{m}-1}e^{s\mathbf{1}({P_{\rm m}\|{\bf x}+{\bf u}_j\|^{-\alpha}>P_{\rm s}\|{\bf u}_j\|^{-\alpha}})}|{\bf x}\bigg]\\
&=\prod\limits_{j=1}^{\bar{m}-1}\nbbE[e^{s\mathbf{1}({P_{\rm m}\|{\bf x}+{\bf u}_j\|^{-\alpha}>P_{\rm s}\|{\bf u}_j\|^{-\alpha}})}|{\bf x}]\notag\\&
= \prod\limits_{j=1}^{\bar{m}-1}e^s\nbbP({P_{\rm m}\|{\bf x}+{\bf u}_j\|^{-\alpha}>P_{\rm s}\|{\bf u}_j\|^{-\alpha}}|{\bf x})\\&+\nbbP(P_{\rm s}\|{\bf u}_j\|^{-\alpha}>P_{\rm m}\|{\bf x}+{\bf u}_j\|^{-\alpha}|{\bf x})\notag\\& = ({\cal A_{\rm m}}(x)e^s + (1-{\cal A_{\rm m}}(x)))^{\bar{m}-1},\notag
\end{align*} 
which is the MGF of a Binomial distribution with $(\bar{m}-1,{\cal A}_{\rm m}(x))$. Here,  the first step follows from the fact that ${\bf u}_j$-s are i.i.d. Similarly for \case~$2$, 
\begin{align*}
&\nbbE[e^{sN_{\bf x}^{\rm ABS,o}}|{\bf x}] =\nbbE\bigg[\prod\limits_{j=1}^{{N}_{{\bf x}}}\nbbE[e^{s\mathbf{1}({P_{\rm m}\|{\bf x}+{\bf u}_j\|^{-\alpha}>P_{\rm s}\|{\bf u}_j\|^{-\alpha}})}|{\bf x}]\bigg]\\
&=\nbbE[ ({\cal A_{\rm m}}(x)e^s + (1-{\cal A_{\rm m}}(x)))^{{N}_{\bf x}}]\\
&=\sum_{k=0}^{\infty}({\cal A_{\rm m}}(x)e^s + (1-{\cal A_{\rm m}}(x)))^{k}\frac{\bar{m}^{k}e^{-\bar{m}}}{k!}=
e^{\bar{m}{\cal A}_{\rm m}(x)(e^{s}-1)},
\end{align*} 
which is the MGF of a Poisson distribution with mean $\bar{m}{\cal A}_{\rm m}(x)$. From the PMF of $N_{\bf x}^{\rm ABS,o}$, one can easily obtain the PMF of $N_{\bf x}^{\rm ABS}$. 
 The PMF of $N_{\bf x}^{\rm SBS}$ can be obtained on similar lines by altering the inequality in the first  step of the above derivation.
\subsection{Proof of Lemma~\ref{lemm::load::characterization::others}}
\label{app::load::characterization::others}
Following the proof of Lemma~\ref{lemm::load::characterization::abs}, conditioned on the location of a hotspot at ${\bf x}_i$, $N_{{\bf x}_i}^{\rm ABS}$ becomes (i) \case~$1$. a Binomial random variable with $(\bar{m}, {{\cal A}}_{\rm m}({ x}_i))$, or (ii) \case~$2$. a Poisson random variable with $\bar{m}{\cal A}_{\rm m}(x_i)$. Now, $N_{\rm o}^{\rm ABS } = \sum_{i=1}^{n-1} \nbbE_{{\bf x}_i}[N_{{\bf x}_i}^{\rm ABS }]$, where $\nbbE_{{\bf x}_i}[N_{{\bf x}_i}^{\rm ABS }]$-s are i.i.d. with $\nbbP(\nbbE_{{\bf x}_i}[N_{{\bf x}_i}^{\rm ABS }]=k) = $
 \begin{align*}
\begin{cases}\int_0^{R-R_{\rm s}}{\bar{m}\choose k}{\cal A}_{\rm m}(x_i)^{k}{\cal A}_{\rm s}(x_i)^{\bar{m}-k}f_{X}(x_i){\rm d}{x_i},&\text{for \case~$1$}\\
\int_{0}^{R-R_{\rm s}}\frac{e^{-\bar{m}{\cal A}_{\rm m}(x_i)}(\bar{m}{\cal A}_{\rm m}(x_i))^k}{k!}f_X(x_i){\rm d}x_i,
&\text{for \case~$2$}\
\end{cases},
\end{align*}where $k\in\nbbZ^+$. The exact PMF of $N_{\rm o}^{\rm ABS }$ is obtained by the $(n-1)$-fold discrete convolution of this PMF. We avoid this complexity of the exact analysis by first characterizing the mean and variance of $N_{\rm o}^{\rm ABS}$ as: $\upsilon_{\rm m}= \nbbE[N_{\rm o}^{\rm ABS}] = \sum_{i=1}^{n-1}\nbbE [\nbbE_{{\bf x}_i}[N_{{\bf x}_i}^{\rm ABS }]]=(n-1)\bar{m}{\cal A}_{\rm m}(X)$, and $\sigma_{\rm m}^2 =$
\begin{align*}
& {\rm Var}[N_{\rm o}^{\rm ABS}]\myeq{a}\sum_{i=1}^{n-1} {\rm Var}[\nbbE_{{\bf x}_i}[N_{{\bf x}_i}^{\rm ABS}]]\\
&=\sum_{i=1}^{n-1} \nbbE[(\nbbE_{{\bf x}_i}[N_{{\bf x}_i}^{\rm ABS}])^2]-(\nbbE[\nbbE_{{\bf x}_i}[N_{{\bf x}_i}^{\rm ABS}]])^2\\
&= \begin{cases}\sum_{i=1}^{n-1}\int_0^{R-R_{\rm s}}(\bar{m}{\cal A}_{\rm m}(x_i){\cal A}_{\rm s}(x_i)+(\bar{m}{\cal A}_{\rm m}(x_i))^2)&\\f_X(x_i){\rm d}{x_i}  -(\bar{m}\nbbE[{\cal A}_{\rm m}(X)])^2, \text{ for \case~$1$}\\
\sum_{i=1}^{n-1} \int_0^{R-R_{\rm s}}(\bar{m}{\cal A}_{\rm m}(x_i)+(\bar{m}{\cal A}_{\rm m}(x_i))^2)f_X(x_i){\rm d}x_i\\ -(\bar{m}\nbbE[{\cal A}_{\rm m}(X)])^2,  \text{ for \case~$2$}
\end{cases},
\end{align*} where (a) is due to the fact that $\nbbE_{{\bf x}_i}[N_{{\bf x}_i}^{\rm ABS }]$-s are i.i.d. The final result follows from some algebraic manipulation. Having derived the mean and variance of  $N_{\rm o}^{\rm ABS}$, we invoke CLT to approximate the distribution of $N_{\rm o}^{\rm ABS}$ since it can be represented as a sum of i.i.d. random varables with finite mean and variance. Similar steps can be followed for the distribution of   $N_{\rm o}^{\rm SBS}$. 
\subsection{Proof of Theorem~\ref{thm::rate::cov::equal::partition}}
\label{app::rate::cov::equal::partition}
First we evaluate $\pr_{\rm m}=\pr_{\rm m} = \nbbP({\cal R}_{\rm a}^{\rm ABS}>\rho)=$ 
\begin{align*}&
\nbbP\bigg(\frac{W_{\rm a}}{N_{\bf x}^{\rm ABS}+N_{\rm o}^{\rm ABS}}\log_2(1+\snr_{\rm a}^{\rm ABS}({\bf x}+{\bf u}))>\rho\bigg)\\&
=\nbbP\bigg(\snr_{\rm a}^{\rm ABS}({\bf x}+{\bf u})>2^{\frac{\rho(N_{\bf x}^{\rm ABS}+N_{\rm o}^{\rm ABS})}{W_{\rm a}}}-1\bigg)\\&= \pc_{\rm m}{\bigg(2^{\frac{\rho(N_{\bf x}^{\rm ABS}+N_{\rm o}^{\rm ABS})}{W_{\rm a}}}-1\bigg)},
\end{align*}
where,  the first step follows from \eqref{eq::rate_abs_access}. The final form is obtained by deconditioning with respect to $N_{\bf x}^{\rm ABS}$, $N_{\rm o}^{\rm ABS}$ and $\bf x$.
 Now for equal partition,  $\pr_{\rm s}=\nbbP({\cal R}_{\rm a}^{\rm SBS}>\rho)=$ %The detailed proof is omitted due to space constraints. 
\begin{align}
&\nbbP\bigg(\frac{W_{\rm b} }{N_{\bf x}^{\rm SBS}n}\log_2(1+\snr_{\rm b}({\bf x}))>\rho\bigg)\notag\\
&\qquad\times\nbbP\bigg(
\frac{ W_{\rm a}}{N_{\bf x}^{\rm SBS}}\log_2(1+\snr_{\rm a}^{\rm SBS}({\bf u}))>\rho\bigg)\notag\\
&=\nbbP\bigg(\snr_{\rm b}({\bf x})>2^{\frac{\rho n N_{\bf x}^{\rm SBS}}{W_{\rm b} }}-1\bigg)\nbbP\bigg(
 \snr_{\rm a}^{\rm SBS}({\bf u})>2^{\frac{\rho N_{\bf x}^{\rm SBS}}{W_{\rm a}}}-1\bigg)\notag\\&=\nbbE\bigg[\pc_{\rm s}\bigg(2^{\frac{\rho n N_{\bf x}^{\rm SBS}}{W_{\rm b} }}-1,2^{\frac{\rho N_{\bf x}^{\rm SBS}}{W_{\rm a}}}-1\big|x\bigg)\bigg].\notag
\end{align}
Here step (a) follows from \eqref{eq::rate_sbs_access} and the fact that the two rate terms appearing under the $\min$ operator are independent. The final form is obtained by deconditioning with respect to $N_{\bf x}^{\rm SBS}$ and $x$.  %(using Lemmas~\ref{lemm::load::characterization::abs} and \ref{lemm::load::characterization::others}). 
For instantaneous load-based partition,  % \vspace{-1em}
\begin{multline}
\pr_{\rm s} =\nbbP\bigg(\frac{W_{\rm b} }{N_{\bf x}^{\rm SBS}+N_{\rm o}^{\rm SBS}}\log_2(1+\snr_{\rm b}({\bf x}))>\rho\bigg)\\\times \nbbP\bigg(
\frac{ W_{\rm a}}{N_{\bf x}^{\rm SBS}}\log_2(1+\snr^{\rm SBS}_{\rm a}({\bf u}))>\rho\bigg)\\=\nbbP\bigg(\snr_{\rm b}({\bf x})>2^{\frac{\rho(N_{\bf x}^{\rm SBS}+N_{\rm o}^{\rm SBS})}{W_{\rm b} }}-1\bigg)\\\times \nbbP\bigg(
 \snr_{\rm a}^{\rm SBS}({\bf u})>2^{\frac{\rho N_{\bf x}^{\rm SBS}}{W_{\rm a}}}-1\bigg)\\
 =\nbbE\bigg[\pc_{\rm s}\bigg(2^{\frac{\rho(N_{\bf x}^{\rm SBS}+N_{\rm o}^{\rm SBS})}{W_{\rm b} }}-1,2^{\frac{\rho N_{\bf x}^{\rm SBS}}{W_{\rm a}}}-1\big|x\bigg)\bigg].\notag
 \end{multline}
 The final form is obtained by deconditioning with respect to $N_{\bf x}^{\rm SBS}$, $N_{\rm o}^{\rm SBS}$ and $x$. For average load-based partition, $\pr_{\rm s}=$
 \begin{multline*}
\nbbP\bigg(\frac{W_{\rm b}\nbbE[N_{\bf x}^{\rm SBS}] }{N_{\bf x}^{\rm SBS}(\nbbE[N_{\bf x}^{\rm SBS}]+\sum_{i=1}^{n-1}\nbbE[N_{{\bf x}_i}^{\rm SBS}])}\log_2(1+\snr_{\rm b}({\bf x}))\\>\rho\bigg)\nbbP\bigg(
\frac{ W_{\rm a}}{N_{\bf x}^{\rm SBS}}\log_2(1+\snr_{\rm a}^{\rm SBS}({\bf u}))>\rho\bigg)\notag\\
=\nbbP\bigg(\frac{W_{\rm b}\nbbE[N_{\bf x}^{\rm SBS}] }{N_{\bf x}^{\rm SBS}(\nbbE[N_{\bf x}^{\rm SBS}]+\bar{m}\sum_{i=1}^{n-1}{\cal A}_{\rm s}(x_i))}\log_2(1+\snr_{\rm b}({\bf x}))\\>\rho\bigg)\nbbP\bigg(\frac{ W_{\rm a}}{N_{\bf x}^{\rm SBS}}\log_2(1+\snr_{\rm a}^{\rm SBS}({\bf u}))>\rho\bigg),\notag%\label{ref::intermediate::rate::coverage:avg::load::based}
 \end{multline*}%\vspace{-1em}
where the last step is obtained by using the fact that $\nbbE[N_{\bf x_i}^{\rm SBS}]=\bar{m}\nbbE[{\cal A}_{\rm s}(X)]$ and $\nbbE[N_{\bf x}^{\rm SBS}]$ depends on the underlying distribution of $N_{\bf x}$. Since the random variable $\sum_{i=1}^{n-1}{\cal A}_{\rm s}(x_i)$ is a summation of $n-1$ i.i.d. random variables with mean $(n-1)\nbbE[{\cal A}_{\rm s}(X)]$ and variance $(n-1){\rm Var}[{\cal A}_{\rm s}(X)]$, we again invoke CLT instead of resorting to the exact expression {that would have involved the} $(n-1)$-fold convolution of the PDF of ${\cal A}_{\rm s}(X)$. 
%{%\setstretch{1.20}
%\bibliographystyle{IEEEtran}
%\bibliography{mmwave_icc,MA_ref,new_ref}
%}

{
\bibliographystyle{IEEEtran}
\bibliography{JournalDraft.bbl}
}

\end{document}